\pdfoutput=1                    %% Need this for accurate compiling of tex files when submitting in ArXiv. %%
\documentclass[twoside,11pt]{article}

\usepackage{fullpage}
\usepackage[authoryear]{natbib}
\usepackage{amsthm, amsfonts, amssymb, amsmath, bbm}
\usepackage{enumerate}  %Replacing the enumitem package by the enumerate package which provides more flexibilities.
\usepackage{mathtools}
\usepackage{amsmath}
\usepackage{bm}
\usepackage{amsfonts}
\usepackage{amssymb}
\usepackage{mathtools}
\usepackage{comment}
\usepackage{enumerate}
\usepackage{multirow}
\usepackage{float}
\usepackage{xcolor}
\DeclarePairedDelimiter{\ceil}{\lceil}{\rceil}
\definecolor{navyblue}{rgb}{0.0, 0.0, 0.5}
\definecolor{bluegreen}{rgb}{0.0, 0.87, 0.87}

\theoremstyle{plain}

\newtheorem{theorem}{Theorem}[section]
\newtheorem{lemma}{Lemma}[section]
\newtheorem{proposition}{Proposition}[section]
\newtheorem{corollary}{Corollary}[section]
\theoremstyle{definition}

\newtheorem{assumption}{Assumption}[section]
\newtheorem{remark}{Remark}[section]

\newcounter{myalgctr}
\numberwithin{myalgctr}{section}

\numberwithin{equation}{section}

\def\tcm{\textcolor{magenta}}
\def\tcr{\textcolor{red}}

\usepackage[plain,noend]{algorithm2e}

\makeatletter
\renewcommand{\algocf@captiontext}[2]{#1\algocf@typo. \AlCapFnt{}#2} % text of caption
\def\@algocf@capt@plain{top}
\renewcommand{\algocf@makecaption}[2]{%
	\addtolength{\hsize}{\algomargin}%
	\sbox\@tempboxa{\algocf@captiontext{#1}{#2}}%
	\ifdim\wd\@tempboxa >\hsize%     % if caption is longer than a line
	\hskip .5\algomargin%
	\parbox[t]{\hsize}{\algocf@captiontext{#1}{#2}}% then caption is not centered
	\else%
	\global\@minipagefalse%
	\hbox to\hsize{\box\@tempboxa}% else caption is centered
	\fi%
	\addtolength{\hsize}{-\algomargin}%
}
\makeatother

\def\cov{hbox{cov}}
\def\rR{\mathbb{R}}
\def\kK{\mathbb{K}}

\def\I{{\cal I}}
\def\L{{\cal L}}

\def\G{{\cal G}}

\def\Z{{\cal Z}}

\def\boxit#1{\vbox{\hrule\hbox{\vrule\kern6pt  \vbox{\kern6pt#1\kern6pt}\kern6pt\vrule}\hrule}}
\def\rjccomment#1{\vskip 2mm\boxit{\vskip 2mm{\color{black}\bf#1} {\color{blue}\bf -- RJC\vskip 2mm}}\vskip 2mm}

\def\diag{\hbox{diag}}
\def\wh{\widehat}
\def\log{\hbox{log}}

\def\var{\hbox{var}}
\def\cov{\hbox{cov}}

\def\trace{\hbox{trace}}

\def\pr{\P}
\def\wh{\widehat}
\def\trans{^{\rm T}}

\def\b1e{{\mathbf e}}

\newcommand{\bgamma}{\mbox{\boldmath $\gamma$}}
\newcommand{\bzeta}{\mbox{\boldmath $\zeta$}}

\newcommand{\bbeta}{\mbox{\boldmath $\beta$}}

\def\bfd{{\bf d}}

\def\bfM{{\bf M}}

\def\bft{{\bf t}}

\def\bfU{{\bf U}}
\def\bfv{{\bf v}}

\def\trans{^{\rm T}}

\def\b1e{{\mathbf e}}

\def\W{{\mathbf W}}
\def\w{{\mathbf w}}
\def\x{{\mathbf x}}
\def\X{{\mathbf X}}
\def\s{{\mathbf s}}
\def\S{{\mathbf S}}
\def\P{{\mathbb P}}

\def\Z{{\mathbf Z}}

\renewcommand{\hat}{\widehat}
\renewcommand{\tilde}{\widetilde}

\newcommand{\hl}{\hat{\ell}_{n,k}}
\def\hm{\hat{m}_{n,k}}

\newcommand{\mnk}{m_{n,k}}
\newcommand{\hf}{\hat{f}_n}

\newcommand{\hphi}{\hat{\phi}_n}
\newcommand{\hphik}{\hat{\phi}_{n,k}}

\newcommand{\vt}{\theta_0}

\newcommand{\bfeta}{{\boldsymbol\eta}}

\newcommand{\hvt}{\hat{\theta}}
\newcommand{\hopt}{h_{\mbox{\tiny opt}}}
\newcommand{\hvts}{\hvt_{\mbox{\tiny SUP}}}
\newcommand{\hvtss}{\hvt_{\mbox{\tiny SS}}}
\newcommand{\hvtgs}{\hvt_{\mbox{\tiny GS}}}

\newcommand{\vti}{\theta_{\mbox{\tiny INIT}}}
\newcommand{\vtos}{\theta_{\mbox{\tiny OS}}}
\newcommand{\hvti}{\hvt_{\mbox{\tiny INIT}}}
\newcommand{\sigsup}{\sigma_{\mbox{\tiny SUP}}}
\newcommand{\sigss}{\sigma_{\mbox{\tiny SS}}}
\newcommand{\hsigss}{\hat{\sigma}_{\mbox{\tiny SS}}}
\newcommand{\sigeff}{\sigma_{\mbox{\tiny EFF}}}

\newcommand{\tsigss}{\tilde{\sigma}_{\mbox{\tiny SS}}}

\newcommand{\tvt}{\tilde{\theta}}

\newcommand{\hpsi}{\hat{\psi}_{n,k}}

\newcommand{\nn}{n^{-1}}

\newcommand{\hga}{\wh{\bgamma}}
\newcommand{\hg}{\wh{g}}

\newcommand{\mbG}{\mathbb{G}}
\newcommand{\mbP}{\mathbf{P}}
\newcommand{\hmbP}{\hat{\mathbf{P}}_k}
\newcommand{\E}{\mathbb{E}}

\def\ms{\mathcal{S}}
\def\mp{\mathcal{P}}
\def\md{ {\cal D}}
\def\mx{ {\cal X}}
\def\mb{ \mathcal{B}}

\def\cl{ \mathcal{L}}
\def\cu{ \mathcal{U}}
\def\ct{ \mathcal{T}}

\def\sm{\hbox{$\sum_{m=1}^{M}$}}
\def\sl{\hbox{$\sum_{i=1}^n$}}
\def\slk{\hbox{$\sum_{i\in\I_{k}^-}$}}

\def\sk{\hbox{$\sum_{k=1}^\kK$}}

\def\bze{\mathbf{0}}
\def\bon{\mathbf{1}}
\def\mbtv{\mb(\vt,\varepsilon)}

\def\sg{\hbox{$\sup_{\bgamma\in\ct}$}}

\def\sgx{\hbox{$\sup_{\x\in\mx,\,\bgamma\in\ct}$}}

\def\sb{\hbox{$\sup_{\theta\in\mbtv}$}}

\def\sbx{\hbox{$\sup_{\x\in\mx,\,\theta\in\mbtv}$}}
\def\sx{\hbox{$\sup_{\x\in\mx}$}}
\def\sxx{\hbox{$\sup_{\x,\X\in\mx}$}}

\def\sups{\hbox{$\sup_{\s\in\ms}$}}

\def\sss{\hbox{$\sup_{\s,\,\S\in\ms}$}}

\def\hL{\hat{L}_n}

\def\mn{\hbox{{\it Normal}}\,}
\def\Enk{\E_{n,k}}
\newcommand{\bmu}{\bfM}

\makeatletter
\def\trans{^{\rm T}}

\def\boxit#1{\vbox{\hrule\hbox{\vrule\kern6pt  \vbox{\kern6pt#1\kern6pt}\kern6pt\vrule}\hrule}}
\def\rjccomment#1{\vskip 2mm\boxit{\vskip 2mm{\color{black}\bf#1} {\color{blue}\bf -- RJC\vskip 2mm}}\vskip 2mm}
\def\rcom#1 {{\color{red}\bf#1} }
\def\bcom#1 {{\color{blue}\bf#1} }
\def\daicomment#1{\vskip 2mm\boxit{\vskip 2mm{\color{black}\bf#1} {\color{blue}\bf -- Dai\vskip 2mm}}\vskip 2mm}
\def\bse{\begin{eqnarray*}}
	\def\ese{\end{eqnarray*}}
\def\be{\begin{eqnarray}}
	\def\ee{\end{eqnarray}}
\newlength{\gnat}
\baselineskip=\gnat

\def\tcr{\textcolor{red}}
\def\tcm{\textcolor{magenta}}
\def\tcb{\textcolor{blue}}

\def\tcrold{\textcolor{red}\bf }   %% Introducing (on 10/22/2021) \tcrold - a separate red text command to indicate/track all my prior edits (before this round) - helps to keep track of things better. Can also set this to black independently, and only keep my new edits highlighted in color - AC.

\def\tcmAC{\textcolor{cyan}} %% Introducing (on 1/16/22) \tcmAC - a separate magenta command to be able to distinguish my own edits within an already magenta colored edited portion. Can also set this to red or black independently -- AC.

\def\tcr{\textcolor{black}}   % Uncomment these to set all commands to black once edits are done - AC.
\def\tcb{\textcolor{black}}

\def\tcm{\textcolor{black}}
\def\tcmAC{\textcolor{black}}

\def\tcrold{\textcolor{black}}

\def\Xtilde{\tilde{\X}}
\def\bbetahatsup{\hat{\bbeta}_{\mbox{\tiny SUP}}}
\def\bbetahatss{\hat{\bbeta}_{\mbox{\tiny SS}}}
\def\bbetahatinit{\hat{\bbeta}_{\mbox{\tiny INIT}}}
\def\Q{Q}

\def\abhishek#1{\vskip 2mm\boxit{\vskip 2mm{\color{red}\bf#1} {\color{red}\bf -- AC\vskip 2mm}}\vskip 2mm}

\usepackage{xr-hyper}
\RequirePackage[colorlinks,citecolor=blue,urlcolor=blue,linkcolor=blue]{hyperref}
\makeatletter
\def\namedlabel#1#2{\begingroup
    #2%
    \def\@currentlabel{#2}%
    \phantomsection\label{#1}\endgroup
}
\makeatother
\newcommand{\vertiii}[1]{{\vert\kern-0.25ex\vert\kern-0.25ex\vert #1
    \vert\kern-0.25ex\vert\kern-0.25ex\vert}}

\usepackage[toc,page]{appendix}

\usepackage{abstract}

\title{\bf \tcmAC{Semi-Supervised Quantile Estimation: Robust and Efficient Inference in High Dimensional Settings}}

\author{
    Abhishek Chakrabortty\thanks{Department of Statistics, Texas A\&M University \tcmAC{(email: \href{mailto:abhishek@stat.tamu.edu}{abhishek@stat.tamu.edu}).}} %. (Email: {\tt abhishek@stat.tamu.edu})}
    \and
    Guorong Dai\thanks{Department of Statistics and Data Science, Fudan University \tcmAC{(email: \href{mailto:guorongdai@fudan.edu.cn}{guorongdai@fudan.edu.cn}).} Guorong Dai is a joint first author. Guorong Dai is the corresponding author.}
    \and
    Raymond J. Carroll\thanks{Department of Statistics, Texas A\&M University \tcmAC{(email: \href{mailto:carroll@stat.tamu.edu}{carroll@stat.tamu.edu}).}} %. (Email: {\tt carroll@stat.tamu.edu})}
}

\begin{document}
 % \editor{}

%\date{}
\maketitle

\vspace{-0.25in}   %Needed to do this to take care of a weird formatting issue - due to the long abstract, the footnote was carrying forward to next page!

\vspace{-0.4in}   %% Adding this for better formatting -- AC (1/22/2022).
\begin{abstract}
\vspace{-0.5in}    %% Changing this for better formatting -- AC (1/22/2022).
\vspace{-0.15in}
We consider quantile estimation in a semi-supervised setting, where \tcr{one has} %there are
two available data sets: (i)  a small or moderate \tcr{sized} \textit{labeled data \tcm{set}} containing observations \tcrold{for} %of
a response and a set of possibly high dimensional covariates, %%possibly high dimensional covariates,
and (ii) a much larger \textit{unlabeled data \tcm{set}} %\tcb{**``Labeled data'' and ``unlabeled data'' are terms in semi-supervised learning.**}
where only the covariates are observed.
\tcb{Such settings are of increasing relevance in modern studies involving large databases where labeled data may be limited due to practical constraints but unlabeled data \tcm{are} plentiful, \tcr{and it is of interest to investigate how the latter may be exploited.}}
%
%\tcb{**Perhaps we can delete this sentence about background.** \tcr{**No need -- this part is important. Plus I have shortened the abstract to fit in one page any way.**}}.
%{\color{magenta}\bf Maybe some references here} \tcb{**References in the abstract are not preferred by Biometrika. The template says ``The summary contains bibliographic references only if they are essential''.** \tcr{**Plus we also discuss references in detail in the Introduction.**}}
%
We propose a family of semi-supervised estimators \tcrold{for the response quantile(s)} based on the two data sets, to improve the estimation accuracy compared to the supervised \tcrold{estimator}, \tcm{i.e., %method
	the sample quantile,} which uses the labeled data only. \tcrold{These estimators are based on a flexible imputation strategy applied to the estimating equation along with a debiasing step that allows for full robustness against misspecification of the imputation model. Further,} \tcrold{a} %The
one-step update strategy is adopted to \tcrold{enable easy implementation of our method and} handle \tcrold{the inevitable} complexity arising from the non\tcr{-}linear nature of the quantile estimating equation. %\tcb{Also,} \tcb{cross-fitting} \tcrold{is used to allow high dimensional imputation models.}
Under \tcrold{fairly mild} %reasonable
assumptions, we prove our estimators are \tcrold{\emph{fully robust}} to the choice of the nuisance imputation model, in the sense of {\it always} maintaining root-$n$ consistency \tcrold{and} asymptotic normality\tcrold{, while having} \tcr{\emph{improved}} efficiency relative to the supervised estimator\tcm{.}
%\emph{always} root-$n$-consistent \tcb{and} asymptotically normal, \tcrold{and improved (and \emph{never} worse than) the supervised estimator with respect to efficiency},
%while
%\tcrold{and also} attaining
%\tcmAC{Further, they also attain}  %% NOTE: Slightly rewording this phrase here to ensure better formatting (in the Arxiv draft only) -- AC (1/22/2022). %%
\tcmAC{Further, they achieve}
semi\tcr{-}parametric \tcr{\emph{optimality}} \tcmAC{also,} \tcrold{provided} %as long as
the relation between the response and the covariates is correctly specified \tcrold{via the imputation model}.
% and first-order insensitive to estimation errors of the nuisance functions
% and no worse than the supervised estimator with respect to efficiency, while attaining semi\tcrold{-}parametric optimality as long as the relation between the response and the covariates is correctly specified.
%\tcrold{Moreover, the estimators are fully robust to the choice of the nuisance imputation model used, always maintaining root-$n$ consistency and improved efficiency, and are also} first-order insensitive to estimation errors of the \tcrold{imputation function.} %nuisance functions
%\tcb{Moreover, our estimators are also first-order insensitive to {\it any} estimation errors of the imputation function.} %nuisance functions
%Moreover,
\tcrold{In addition,} as an \tcrold{illustration of estimating} %instance of approximating
the nuisance imputation function, %s \tcrold{required} in our method,
%\tcrold{i.e., the imputation function, required in our method,}
\tcrold{we consider kernel smoothing type estimators on lower dimensional and possibly estimated transformations of the high dimensional covariates, and we} establish novel results on %the
\tcr{\emph{uniform}} convergence rate\tcrold{s} of \tcrold{such} kernel smoothing estimators \tcr{in \emph{high dimensions},} \tcrold{involving responses indexed by a function class} %{with a function class of a response, high dimensional covariates and dimension reduction techniques involved}
\tcrold{and usage of dimension reduction techniques.} %on the high dimensional covariates}.
\tcrold{These results may be of independent interest.} %{Numerically, outcomes of}
\tcrold{Numerical results}
on both simulated and real data confirm our semi-supervised approach\tcr{'s improved performance, both in terms of estimation as well as inference.} %attains better estimation efficiency than its supervised counterpart while \tcr{also} working satisfactorily in terms of inference. %\tcr{thus validating our theory  in finite samples}. %\tcrold{and also maintaining full root-$n$ rate robustness against the imputation used}.

% \abhishek{Abstract corrections not finalized yet. Will work on this later.}
\end{abstract}
\par\smallskip

%\begin{keywords}
\noindent
{\bf Keywords:} \tcm{Semi-supervised \tcmAC{inference, Q}uantile estimation, \tcmAC{Robustness and efficiency,} %% NOTE: Changing some of the keywords to make them informative and distinct from the new title (and avoid direct overlap between the keywords and phrases appearing in the new title). Doing this  in both Arxiv and Overleaf drafts -- AC (1/25/2022) %Semi-supervised quantile estimation, Robust and efficient inference,
Imputation and debiasing, High dimensional nuisance estimator\tcr{s}, \tcmAC{Kernel smoothing \tcr{with dimension reduction}.}} %Sample splitting.} %and cross-fitting.}  %NOTE: I am removing this keyword entirely. Don't think it's that important -- AC (1/22/2022).
\pagebreak\newpage  %NOTE: Removing this for the Arxiv draft as of 1/22/2022 -- AC. %%

\section{Introduction}\label{seci}
%\subsection{Background and literature review}
%%
%Semi-supervised learning has been attracting increasing attention as an exciting new area in statistics and machine learning.
%Typically, a semi-supervised setting is
\tcr{Semi-supervised settings, as the name suggests, are} characterized by two available data sets: (i) a {\it labeled data} containing a limited number of observations on both a response $Y$ and a set of covariates $\X$, and (ii) an {\it unlabeled data} of much larger size where only the covariates $\X$ are observed. \tcr{Statistical learning in such settings, often termed ``semi-supervised learning'', has gained substantial attention in the last two decades in machine learning, and more recently, in the statistics community as well.} A detailed \tcr{overview of semi-supervised learning and  %survey on
	the growing literature} %and recent developments of this topic
can be found in \citet{zhu2009introduction} and \citet{chapelle2010semi}.
%%%
%Semi-supervised learning has \sout{been attracting increasing attention}
%\tcrold{emerged} as a promising area of \tcrold{research in modern} statistics and machine learning \sout{in recent years}.
%A detailed \sout{survey on}
%\tcrold{overview of}
%the growing literature and recent developments \tcrold{on} \sout{of}
%this topic can be found in \citet{zhu2009introduction} and \citet{chapelle2010semi}, \sout{while Chapter 2 of \citet{Chakrabortty_Thesis_2016} provides an elaborate interpretation from a semi\tcr{-}parametric perspective.}
Semi-supervised settings emerge naturally when observations \tcr{for} %of
$\X$ are easy to collect for a large cohort\tcmAC{,} but \tcr{obtaining the corresponding} %the identification process of
$Y$ is costly or time-consuming.
This situation is \tcr{ubiquitous in modern studies across scientific disciplines (see \citet{chapelle2010semi} for various examples), including modern biomedical studies,} such as   %is commonplace in, for example,
electronic health records \tcr{--} rich resources of data for discovery research \tcr{--} where labeling of $Y$ is often logistically prohibitive; see Section 1 of \citet{chakrabortty2018efficient} for further details. Another highly relevant biomedical application is \tcmAC{in} integrative genomics, especially expression quantitative trait loci studies \citep{michaelson2009detection} that develop association mapping between gene expression levels $Y$ and genetic variants $\X$.  A major bottleneck for the power of such studies has turned out to be the limited sample size of \tcr{the} expensive gene expression data \citep{flutre2013statistical, mccaw2021cross}. In contrast, observations \tcr{for} %of
the genetic variants are usually cheaper and available for much more individuals. Safely and efficiently leveraging these plentiful unlabeled data \tcr{thus} necessitates \tcr{developing} suitable semi-supervised strategies for robust and efficient inference, which are our main focus in this paper.

%a rich source of data  to drive discovery in disease genomics \citep{kohane2011using}. In statistical analysis based on electronic medical records, labels of phenotype information are usually available only in a small fraction of a large cohort, since there is no pre-specified outcome in the data collection while the identification process can be costly and time-consuming \citep{cheng2020robust}. In contrast, values of clinical variables have been recorded digitally for each individual. Leveraging these plentiful data without labels safely and efficiently necessitates suitable semi-supervised strategies.

%\vspace{-0.05in}
%{\it \tcr{Our goal} and existing work.}
\subsection{\tcm{\tcr{Our goal} and existing \tcmAC{work}}}  %.}} %% No period here in the subsection title -- AC 1/22/2022.
In semi-supervised settings, the most critical \tcm{and natural} question is when and how \tcr{the} unlabeled data can be used to improve estimation accuracy compared to {\it supervised} methods, which take account of \tcmAC{the}
labeled data only. In principle, such improvement is determined by how the parameter of interest depends on the marginal distribution $\P_\X$ of $\X$ as the unlabeled data are informative for $\P_\X$ only \citep{seeger2000learning, zhang2000value}. Thus conditions on the relation between $\P_\X$ and the conditional distribution of $Y$ given $\X$ were assumed \tcm{explicitly or implicitly} by many existing semi-supervised approaches \tcr{that aimed} %aiming
to estimate $\E(Y\mid\X)$
%\tcmAC{parametrically or}
nonparametrically, including generative modeling \tcr{\citep{nigam2000text},}  %\citep{nigam2000text, nigam2001using},
graph-based methods \citep{zhu2005semi} and manifold regularization \citep{belkin2006manifold}\tcr{. %, whose advantage
	Their advantages} over supervised learning\tcr{, however,} cannot be guaranteed when the\tcr{se} distribution\tcr{al} assumptions are violated \citep{cozman2001unlabeled, cozman2003semi}. Recently, robust semi-supervised methods were proposed for \tcr{various (finite dimensional) \emph{inference} problems, including} mean estimation \citep{zhang2019semi,zhang2019high}, linear regression \citep{chakrabortty2018efficient, azriel2021semi} and prediction accuracy evaluation \citep{gronsbell2018semia}, among others. As the name ``robust’' suggests, they have been proven to be at least as efficient as their supervised counterparts even if the underlying distribution \tcm{or model assumptions} are misspecified. In a more general context, Chapter 2 of \citet{Chakrabortty_Thesis_2016} provides an elaborate interpretation about robustness and efficiency of semi-supervised inference from a semi\tcr{-}parametric perspective. Despite \tcmAC{this} %the
rich \tcr{and growing} literature \tcr{on semi-supervised inference, the problem of} \tcr{\emph{quantile estimation}} has, to the best of our knowledge, not been investigated in semi-supervised settings. \tcr{Quantile(s), however, are} %though an
important parameter\tcr{(s) that help characterize} %characterizing
the whole response distribution, and \tcr{are often \tcm{robust} choices for} measuring the central tendency of heavy-tailed \tcr{and/or skewed} data, \tcr{compared to the}  %whose population
mean (\tcr{which} may not \tcr{even} exist). This work therefore aims to bridge this gap in the existing semi-supervised \tcmAC{learning} literature regarding inference for quantiles.

%\vspace{0.05in}
%{\it Problem setup.}
\subsection{\tcm{Problem setup}}
To formulate our work, suppose $\X\in\mx\subset\rR^p$, \tcm{ for some $\mx$,} and $Y \tcr{\in \rR}$
is continuous with a density function $f(\cdot)$ as well as a distribution function $F(\cdot)$.
Here, the dimension $p\equiv p_n$ of $\X$  could \tcr{\emph{diverge},} and \tcr{possibly} exceed \tcr{the sample size} $n$, allowing for the case where $p\gg n$.
%\tcr{Here $\X$ could be \emph{high dimensional}, with} $p \equiv p_n$ \tcr{allowed to} diverge, and \tcr{possibly} exceed \tcr{the (labeled) sample size} $n$, \tcr{including} the case where $p\gg n$.
Our goal is to estimate the \tcr{\emph{$\tau$-quantile},} $\vt\equiv\vt(\tau)\in\Theta\subset\rR$, \tcr{for some $\Theta$}, of the response $Y$, \tcr{for some fixed and known $\tau\in(0,1)$,} defined as the solution \tcr{to} %of
the \tcr{{estimating equation}:}
\be
\E\{\psi(Y,\theta_0)\}~=~0\tcr{,} \label{def}
\ee
%for some fixed and known $\tau\in(0,1)$,
where $\psi(y,\theta):=I(y<\theta)-\tau$ with $I(\cdot)$ \tcr{being} the indicator function. It is important \tcr{to} \tcb{notice} %{**``Note that'' is prohibited by Biometrika.**}
that the \tcr{existing} robust semi-supervised %\tcr{inference}
methods \tcr{for the problems as} mentioned in the last paragraph {\it cannot} be easily extended to this problem, because they all \tcr{directly} impute \tcr{the unlabeled} $Y$ %impute $Y$
with the conditional mean $\E(Y\mid\X)$, or relevant working models \tcr{thereof}.
Due to the \tcr{{non-linear} \tcm{and  {inseparable}}} nature of $Y$ and $\vt$ in the equation \eqref{def}, such \tcr{imputation strategies} %imputations
can, however, lead to \tcr{\emph{bias} in this case.} This fact makes quantile estimation \tcr{fundamentally} %essentially
{different} and %significantly
\tcr{much more} {challenging} compared to problems based on estimating equations \tcr{that are linear/separable in the response $Y$ and the parameter(s) of interest,} %with separable responses and parameters of interest,
such as mean estimation \citep{zhang2019semi,zhang2019high} and linear regression \citep{chakrabortty2018efficient, azriel2021semi}.

In a semi-supervised setting, we have two independent data sets: the labeled data $\cl:=\{\Z_i:=(Y_i,\X_i\trans)\trans:i=1,\ldots,n\}$ and the unlabeled data $\cu:=\{\X_i:i=n+1,\ldots,n+N\}$, which
consist of \tcr{$n$ and $N$} independent copies of $\Z:=(Y,\X\trans)\trans$ and $\X$, respectively. %Unlike missing data problems where \tcr{the} missingness \tcr{mechanism} is random,
%\tcr{Here, $\cl$ and $\cu$ are assumed to have the same underlying distribution.}
\tcr{Here, the labeling mechanism is typically by design, \tcm{so that $\cl$ and $\cu$ have the same distribution -- a standard, often impl\tcmAC{i}cit, %% NOTE: noticed this spelling typo just before arxiving. Fixed this in the arxiv draft. However, trying to correct this in the journal version seems to cause major formatting changes. Keeping it unchanged for now (even though there will be a typo) in the journal draft -- AC (1/25/2022).
		assumption in the literature; see \citet[Remark 2.1]{chakrabortty2018efficient} for details --} and the labeling status\tcmAC{, i.e., labeled or not,} of an individual is considered fixed/non-random, \tcm{unlike usual missing data problems where the missingness mechanism/indicator is considered random}.}
%\tcb{being labeled or not here is deterministic for the observations.} \tcb{**perhaps use ``nonrandom'' or ``fixed"**} ({\color{magenta}\bf ????}) \tcr{**I have changed the language above (and added some imp. details too)**}
\tcr{More importantly, a} major \tcr{challenge and a {key feature} \tcm{-- often a consequence of the underlying practical circumstances --} {contrasting} our setup from traditional missing data problems}
%difficulty in our setup
is that the
\tcr{unlabeled data can be \emph{much larger} in size than the labeled data  ($N \gg n$),} %labeled data could be arbitrarily small relative to the unlabeled data,
i.e., it is possible that
\bse
\hbox{$\lim_{n,N\to\infty}$}(n/N)~=~0.
\ese
%\abhishek{Ray, you had some comments here (for the part above) in the earlier version. So I have rewritten this whole part, hoping it's better this time. I haven't kept your comments in the pdf as they don't apply anymore. But you can still check them from the tex file.}
An example is the {\it ideal semi-supervised setting} where $n$ is finite and $N=\infty$, \tcm{i.e., the distribution $\P_\X$ is known}. This clearly \tcr{{violates}} the ``positivity assumption''\tcr{,} that the proportion of $Y$ observed in the whole data $\cl\cup\cu$ is bounded away from zero, often required in the missing data literature \citep{tsiatis2007semiparametric, little2019statistical}. Thus, %for
letting $\nu_{n,N}:=n/(n+N)$, we {allow}
\bse
\nu~:=~\hbox{$\lim_{n,N\to\infty}$}\nu_{n,N}~\in~[0,1).
\ese
Th\tcr{is natural} violation of the positivity assumption caused by the case $\nu=0$ in fact raises substantial technical difficulties, \tcm{e.g., {non-standard} asymptotics}, that cannot be handled by the classical theory of missing data, making our semi-supervised setting fairly unique and challenging.

\subsection{Our contributions}
%\vspace{0.05in}
%{\it Our contributions.}
Based on the labeled data $\cl$ only, a {\it supervised} estimator $\hvts$ can be obtained by solving the sample version of (\ref{def}) given by $\nn\sl\{I(Y_i<\hvts)-\tau\}=0$. Although $\hvts$ possesses consistency and asymptotic normality, which will be shown in Proposition \ref{p1}, ignoring the unlabeled data $\cu$ generally leads to a loss of efficiency; see Remark \ref{remarkeff} for details. To improve the estimation of $\vt$, this article proposes \tcrold{{a family of semi-supervised estimators}}, which takes both $\cl$ and $\cu$ into consideration by imputing the function $\psi(Y,\theta)$ in the definition (\ref{def}). Under mild conditions, our estimators are  \tcrold{\emph{always $n^{1/2}$-consistent for $\vt$ and asymptotically normal}} with arbitrary imputation functions (Theorem \ref{thos}), \tcrold{and also ensured to \emph{outperform}} the supervised estimator $\hvts$ with respect to asymptotic variance (Remark \ref{remarkeff}). When the imputation function correctly specifies the conditional distribution of $Y$ given $\X$, our method further attains  \tcrold{\emph{semi\tcr{-}parametric efficiency}} (Remark \ref{remarkeff}). Another advantage of our estimators is \tcr{their} \emph{first-order insensitivity}, that is, their influence functions are  \tcrold{{not}} affected by estimation errors or %construction
knowledge of \tcr{the} nuisance estimator's \tcr{construction} (Remark \ref{remarkif}). This property is particularly desirable for inference when the covariates $\X$ are  \tcrold{{high dimensional}} or the nuisance functions are \tcr{estimated} %approximated
by nonparametric techniques. \tcrold{Our method uses sample splitting/cross-fitting techniques to allow for such nuisance estimators.} In the proof of these claimed properties, technical innovations related to empirical process theory are established to handle the classes of $\psi(Y,\theta)$ and other relevant random functions indexed by $\theta$; see Lemma \ref{1v2} in Appendix \ref{sm_lemmas}. When constructing our estimators, we adopt the strategy of  \tcrold{{one-step update}} \citep{van2000asymptotic, tsiatis2007semiparametric} to overcome computational difficulties arising from the inseparability of $Y$ and $\theta$ in \eqref{def}, providing a  \tcrold{{simple implementation}} of our estimation and inference procedures. It also avoids the burdensome task of estimating nuisance functions for the whole parameter space $\Theta$. Instead, we only need to consider  \tcrold{{one}} single value of $\theta$. This feature allows us to use a wide range of approaches, including parametric regression \tcrold{and nonparametric/machine learning approaches, like} kernel smoothing and random forest, for the  \tcrold{{nuisance estimation}}, a fairly important component in the implementation of our method that poses substantial challenges in {high dimensional} scenarios. Although the above-mentioned desirable properties of our method are guaranteed by {\it any} nuisance estimators as long as the  \tcrold{{high-level}} conditions in Theorem \ref{thos} hold, we thoroughly study  \tcrold{\emph{kernel smoothing estimators}}, with possible use of  \tcrold{\emph{dimension reduction}}, as an \tcrold{illustration.} %instance.
Specifically, we show those high-level conditions are satisfied by kernel smoothing estimators through deriving their uniform convergence rates, when the outcome involves a function class of $Y$ and the covariates are generated by transforming the high dimensional $\X$ via possibly unknown dimension reduction processes; see Theorem \ref{thhd} and Remark \ref{rkernel}. These results  \tcrold{{extend}} the existing theory of kernel smoothing estimators with \tcrold{{generated covariates}} \citep{mammen2012nonparametric, escanciano2014uniform, mammen_rothe_schienle_2016} %\tcrold{and are also} %while also
%useful in other applications. Hence they should be of independent interest.
\tcr{in {high dimensions}. These are also useful in other applications, and should be of independent interest.}
%
%Summing up, our main contributions thus are:
\tcr{In summary,} our main contributions \tcr{thus} are:
%\vspace{-0.1in}
\begin{enumerate}[(a)]
	\item We develop a \tcrold{{globally robust and locally optimal}} strategy for quantile estimation in semi-supervised and high dimensional setups, %equipping us with % DIfferent from the journal version. -- Dai 01/24/22
	\tcm{providing} a family of $n^{1/2}$-consistent, asymptotically normal and first-order insensitive estimators ensured to be \tcrold{\emph{at least as efficient}} as the sample quantile $\hvts$ and more efficient whenever possible; see Theorem \ref{thos} and Remarks \ref{remarkif}--\ref{remarkeff};
	%\vspace{-0.05in}
	\item As an {\tcrold{illustration} %instance
		of the nuisance estimators \tcr{and their theoretical properties}} required in our method, we consider kernel smoothing \tcr{type} estimators, and derive their uniform convergence rates when the response is indexed by a function class and the high dimensional covariates are transformed by (unknown) dimension reduction mechanisms;
	see Theorem \ref{thhd} and Remark \ref{rkernel}.
\end{enumerate}
%\vspace{-0.1in}

\begin{comment}
	\begin{enumerate}[1)]
		\item We develop a globally robust and locally optimal strategy for quantile estimaiton in semi-supervised and high dimensional setups, equiping us with a family of $n^{1/2}$-consistent, asymptotically normal and first-order insensitive estimators guaranteed to be at least as efficient as the sample quantile $\hvts$ and more efficient whenever possible.
		
		\item We derive the uniform convergence rates of kernel smoothing estimators when the outcome is a function class and dimension reduction techiques are employed to handle high dimension.
	\end{enumerate}
\end{comment}

%\vspace{0.05in}
%{\it Organization of the article.}
\subsection{Organization of the \tcmAC{article}}  %.}} %% No period here in the subsection title -- AC 1/22/2022.
\tcm{In the next section, we introduce our family of semi-supervised estimators for the response quantile $\vt$ and study their asymptotic properties. The choice and estimation of the nuisance functions involved in our approach are theoretically investigated in Section \ref{secnf}. Section \ref{secs} provides numerical results from extensive simulations under a wide range of data generating mechanisms, followed by an empirical data example in Section \ref{secda}. These numerical results substantiate the properties and advantages of our approach stated in the previous sections. Finally, Section \ref{sec_discussion} ends the article with a concluding remark as well as a brief discussion of possible future work. All technical details, including auxiliary lemmas and proofs of all theoretical results, and extra numerical results as well as necessary supplements to the data analysis in Section \ref{secda} can be found in Appendices \ref{sm_technical}--\ref{sm_data_analysis}.}

%%%%########+++++++++++++++++++++++##############%%%%%%%%%%
%%%%########+++++++++++++++++++++++##############%%%%%%%%%%

\section{Semi-supervised estimation of quantiles \tcmAC{via}
	%based on % DIfferent from the journal version. -- Dai 01/24/22
	one-step update}\label{secos}

\paragraph{Notation.}
Throughout, we
use the lower \tcmAC{case} letter $c$ to represent a generic positive constant, including $c_1$, $c_2$, etc., which may vary from line to line. For a $d_1\times d_2$ matrix $\mbP$ whose $(i,j)$th component is $\mbP_{[ij]}$, %let
\tcmAC{we define}
$\hbox{$\|\mbP\|_0:=\max_{1\leq j\leq d_2}\{\sum_{i=1}^{d_1}I(\mbP_{[ij]}\neq 0)\}$, $\|\mbP\|_1:=\max_{1\leq j\leq d_2}(\sum_{i=1}^{d_1}|\mbP_{[ij]}|)$}$,
$\|\mbP\|:=\hbox{$\max_{1\leq j\leq d_2}\{(\sum_{i=1}^{d_1}\mbP_{[ij]}^2)^{1/2}\}$ and $\|\mbP\|_\infty:=\max_{1\leq i\leq d_1, 1\leq j\leq d_2}|\mbP_{[ij]}|$}$.
The symbols $\bon_d$ and $\bze_d$ refer to $d$-dimensional vectors of ones and zeros, respectively \tcmAC{ and $\mn(\mu,\sigma^2)$ denotes the Normal distribution with mean $\mu$ and variance $\sigma^2$}. %% NOTE: Adding this description for Normal in the Arxiv draft (only). Doesn't hurt to clarify it -- AC (1/25/2022).
\tcmAC{We d}enote $\mb(\alpha,\varepsilon):=\{a:|a-\alpha|\leq\varepsilon\}$ as a generic neighborhood of a scalar $\alpha$ with some \tcr{radius} $\varepsilon>0$. For a vector $\bbeta$, we use $\bbeta_{[j]}$ to \tcr{denote} %mean
its $j$th component. For any random function $\hg(\cdot,\theta)$ and \tcmAC{a} random vector $\W$ with copies $\W_1,\ldots,\W_{n+N}$, \tcmAC{we} denote $
\E_{\W}\{\hg(\W,\theta)\}:= \hbox{$\int$} \hg(\w, \theta) dF_{\W}(\w)
$
as the expectation of $\hg(\W,\theta)$ with respect to $\W$ treating $\hg(\cdot,\theta)$ as a non\tcr{-}random function, where $F_{\W}(\cdot)$ is the distribution function of $\W $. For $M\in\{n,n+N\}$, \tcmAC{we} write $\E_M\{\hg(\W,\theta)\}:= M^{-1}\hbox{$\sum_{i=1}^M$} \hg(\W_i,\theta)$ and $\mbG_M\{\hg(\W,\theta)\}:= M^{1/2}[\E_M\{\hg(\W,\theta)\}-\E_\W\{\hg(\W,\theta)\}]$ as well as \tcmAC{define} $\var_M\{\hg(\W,\theta)\}:=\E_M[\{\hg(\W,\theta)\}^2]-[\E_M\{\hg(\W,\theta)\}]^2$. Also,
\tcmAC{we let} $\E_N\{\hg(\W,\theta)\}:= N^{-1}\hbox{$\sum_{i=n+1}^{n+N}$} \hg(\W_i,\theta)$ and $\mbG_N\{\hg(\W,\theta)\}:= N^{1/2}[\E_N\{\hg(\W,\theta)\}-\E_\W\{\hg(\W,\theta)\}]$. Lastly, let $f(\cdot\mid\w)$ and $F(\cdot\mid\w)$ represent the conditional density and distribution functions of $Y$ given $\W=\w$\tcmAC{, respectively}.

\subsection{\tcr{The s}upervised estimator}
We first investigate the supervised estimator $\hvts$\tcmAC{, i.e., the sample quantile, solving:} %solving
$\E_n\{I(Y<\hvts)-\tau\}=0$. To study its limiting behavior, we need the following \tcr{basic} assumption.
\begin{assumption}
	\label{adensity}
	The parameter $\vt$ is in the interior of the space $\Theta$. The density function $f(\cdot)$ of $Y$ satisfies $f(\theta_0)>0$ and has a bounded derivative in $\mbtv$.
\end{assumption}

The basic Assumption \ref{adensity} \tcr{is fairly standard, and it} %that
guarantees the identifiability of $\vt$ \tcmAC{as well as}
%and
the validity of $\hvts$. %Then
\tcmAC{T}he following proposition \tcmAC{then} gives the limiting properties  of $\hvts$.
\begin{proposition}\label{p1}
	Under Assumption \ref{adensity}, the supervised estimator $\hvts$ satisfies\tcmAC{:}
	\bse
	\hvts-\vt~=~-\{nf(\vt)\}^{-1}\sl\psi(Y,\vt)+o_p(n^{-1/2}).
	\ese
	Further\tcmAC{more}, the asymptotic distribution of $\hvts$ is\tcmAC{:}
	\bse
	n^{1/2}f(\vt)\sigsup^{-1}(\hvts-\vt)~\to~ \mn(0,1)\quad (n\to\infty),
	\ese
	with $\sigsup^2:=\var\{\psi(Y,\vt)\}=\tau(1-\tau)$.
\end{proposition}
Proposition \ref{p1} provides the asymptotic variance of $\hvts$, which can be compared with that of our semi-supervised estimator\tcr{(s) proposed} below. Its proof can be found in \citet{k2005}.

\subsection{A family of semi-supervised estimators based on one-step update}\label{sec_one_step}
%\tcr{\emph{Main idea.}}
\paragraph{\tcr{Main idea.}}
The conditional expectation of the left hand side in (\ref{def}) given $\X$ is\tcm{:}
\bse
\tcm{\E\{\psi(Y,\vt)\mid \X\}\equiv F(\vt\mid\X)-\tau\neq 0\hbox{ with a positive probability\tcmAC{,} }}
\ese
if we exclude the trivial situation where $F(\vt\mid\X)=\tau$ almost surely. This indicates \tcmAC{that} the distribution $\P_\X$ of $\X$ \tcmAC{indeed} plays a role in the definition of $\vt$. Hence\tcmAC{,} the supervised estimator $\hvts$ is possibly sub\tcmAC{-}optimal in that it discards the unlabeled data $\cu$, which provides extra information \tcmAC{on} $\P_\X$. To make use of $\cu$ where $Y$ is not observed, we now consider substituting a function of $\X$ for $\E\{\psi(Y,\theta)\}$ in the definition (\ref{def}) of $\vt$. An intuitive choice is $\mu(\X,\theta):=\E\{\psi(Y,\theta)\mid\X\}$ since
$
\E\{\psi(Y,\theta)\}=\E\{\mu(\X,\theta)\}.
$
Then\tcmAC{,} a further representation of $\E\{\psi(Y,\theta)\}$ is\tcm{:}
\bse
\E\{\psi(Y,\theta)\}~=~\E\{\mu(\X,\theta)\}+\E\{\psi(Y,\theta)-\mu(\X,\theta)\}.
\ese
However, the form of $\mu(\X,\theta)$ is typically hard to specify \tcmAC{correctly in practice}. We thus posit a \tcmAC{\it working model} $\phi(\X,\theta)$ %\tcmAC{for it}
\tcr{that is possibly misspecified, and} {\it not} necessarily equal to $\mu(\X,\theta)$.

\tcr{Then,} our method is inspired by the fact that \tcr{the following \emph{robust representation} holds:}
\be
\E\{\psi(Y,\theta)\}~=~h(\theta)~:=~\E\{\phi(\X,\theta)\}+\E\{\psi(Y,\theta)-\phi(\X,\theta)\}\tcmAC{,}
\label{debiase}
\ee
for an \tcr{\emph{arbitrary}} function $\phi(\cdot,\cdot)$, implying \tcr{that under \eqref{def}, at $\theta = \vt$,}
\be
h(\vt)~\equiv~\E\{\phi(\X,\vt)\}+\E\{\psi(Y,\vt)-\phi(\X,\vt)\}~=~0.
\label{popee}
\ee
In the left hand side of (\ref{popee}), the term $\E\{\psi(Y,\vt)-\phi(\X,\vt)\}$ guarantees \tcmAC{\it global robustness} which means the equation always holds\tcmAC{, i.e.,} for \tcmAC{any} function $\phi(\cdot,\cdot)$, while the other term $\E\{\phi(\X,\vt)\}$ involves $\X$ \tcmAC{only} and can thus be estimated using the whole data $\cl\cup\cu$.

\tcmAC{Then}\tcmAC{,} the sample version of (\ref{popee}) constructed with $\cl\cup\cu$ is\tcmAC{:}
\be
\E_{n+N}\{\hphi(\X,\theta)\}+\E_n\{\psi(Y,\theta)-\hphi(\X,\theta)\}~=~0,
\label{samee}
\ee
where $\hphi(\cdot,\cdot)$ denotes \tcmAC{some} reasonable estimator of $\phi(\cdot,\cdot)$ based on $\cl$.

%\vspace{0.05in}
%\tcr{\emph{Construction of the estimators}.}
\paragraph{\tcr{Construction of the estimators.}}
\tcr{Intuitively, the equation \eqref{samee} gives the road map to our semi-supervised estimators.} Solving (\ref{samee}) with respect to $\theta$ is, however, not straightforward owing to its non\tcr{-}linear nature\tcr{. S}o we consider a \tcr{more implementation-friendly and computationally efficient {one-step update}} \tcr{approach}. \tcb{Noticing \tcr{that} the derivative $h'(\cdot)$ of the function $h(\cdot) $ defined in \eqref{debiase} is the density function $f(\cdot)$ of $Y$,  we can hence solve the equation \eqref{popee} by Newton's method, which {refines} an initial solution $\vti$ by a one-step update $\vti-\{h'(\vti)\}^{-1}h(\vti)\equiv$}
\be
\vti+\{f(\vti)\}^{-1}[\E\{\phi(\X,\vti)-\psi(Y,\vti)\}-\E\{\phi(\X,\vti)\}].
\label{one_step_update}
\ee
\tcb{Recall \tcr{from \eqref{samee}} $\hphi(\cdot,\cdot)$ is an estimator of $\phi(\cdot,\cdot)$ based on $\cl$. Further, let $\hvti$ be an \tcm{initial} estimator of $\vt$ and let $\hf(\cdot)$ be an estimator of $f(\cdot)$. Then\tcmAC{,} the empirical version, based on the whole data $\cl\cup\cu$, of the population-level representation \eqref{one_step_update} immediately gives a family of {\it semi-supervised estimators} \tcr{$\hvtss$ of $\vt$} indexed by $\{\hphi(\cdot,\cdot),\hvti,\hf(\cdot)\}$:}
\be
\hvtss~:=~\hvti +\{\hf(\hvti)\}^{-1}[\E_n\{\hphi(\X,\hvti) - \psi(Y,\hvti)\}-\E_{n+N}\{\hphi(\X,\hvti)\}].
\label{defi}
\ee
\tcb{Although we do {not} require specific forms of $\{\hvti,\hf(\cdot)\}$, a natural choice of the initial estimator $\hvti$ is the supervised estimator $\hvts$, while $\hf(\cdot)$ can be a kernel density estimator; see Remark \ref{rcondition} for \tcr{details on} their convergence properties. As regards the imputation function $\hphi(\cdot,\cdot)$, which is an important component in our method, an arbitrary choice is allowed as long as the high-level conditions in Section \ref{sec_main_results} are satisfied. We will thoroughly study some specific examples of $\hphi(\cdot,\cdot)$ in Section \ref{secnf}. However, \tcr{regardless of the choice of $\hphi(\cdot,\cdot)$,} we apply %the general
	\tcr{a} general \emph{cross-fitting} strategy \citep{chernozhukov2018double, newey2018cross} to obtain
	\tcr{$\hphi(\X_i,\cdot)$} as follows. %$\hphi(\cdot,\cdot)$ as follows.
}

%\vspace{0.05in}
%\tcb{\emph{Cross-fitting of $\hphi(\cdot,\cdot)$  and its benefits}.}
\paragraph{\tcb{Cross-fitting of $\hphi(\cdot,\cdot)$  and its benefits.}}
For some fixed integer $\kK\geq 2$, we divide the index set $\I=\{1,\ldots,n\}$ into $\kK$ disjoint subsets $\I_1,\ldots,\I_\kK$ of the same size $n_\kK:=n/\kK$ without loss of generality. Let $\hphik(\cdot,\cdot)$ be \tcr{the corresponding} %an
estimator of $\phi(\cdot,\cdot)$ based on the data $\cl_k^-:=\{\Z_i:i\in\I_k^-\}$ of size $n_{\kK^-}:=n-n_\kK$, where $\I_k^-:=\I/\I_k$. Then we set
\be
\hphi(\X_i,\theta)~\equiv~\sk\{\hphik(\X_i,\theta)I(i\in\I_k)+\kK^{-1}\hphik(\X_i,\theta)I(i>n)\}
\label{ds}
\ee
Through cross-fitting, the dependence of $\hphi(\cdot,\cdot)$ and $\X_i$ in $\hphi(\X_i,\theta)$ $(i=1,\ldots,n)$ is eliminated, so that the second-order errors in the expansion of $\hvtss$ become more tractable while the influence function remains unchanged. We can therefore avoid some stringent conditions, which are analogous to the stochastic equicontinuity ones in the empirical process theory \citep{van2000asymptotic}, when deriving properties of $\hvtss$. More detailed discussions of cross-fitting can be found in \citet{chakrabortty2018efficient}, \citet{chernozhukov2018double} and \citet{newey2018cross}.

%The motivation for the cross-fitting is to bypass technical challenges arising from the dependence of $\hphi(\cdot,\cdot)$ and $\X_i$ in the term $\hphi(\X_i,\theta)$ $(i=1,\ldots,n)$. Without cross-fitting, the same theoretical conclusions require more stringent assumptions analogous to the stochastic equicontinuity conditions in the classical theory of empirical process\tcr{es \citep{van2000asymptotic}}. These assumptions are typically hard to verify and less likely to hold in high dimensional scenarios. Essentially, using cross-fitting makes the second-order errors in the stochastic expansion of $\hvtss$ easier to control while not changing the first-order properties, i.e., the influence function of $\hvtss$. See Theorem 4.2 and the following discussion in \citet{chakrabortty2018efficient}, as well as \citet{chernozhukov2018double} and \citet{newey2018cross}, for further insights concerning cross-fitting.

%based on its ``influence function'' $\{f(\vt)\}^{-1}[\phi(\X,\vt)-\psi(Y,\vt)-\E_{n+N}\{\phi(\X,\vt)\}] $\tcr{.} Let $\hvti$ and $\hf(\cdot)$ be an initial estimator of $\vt$ and an estimator of the density function $f(\cdot)$, both using $\cl$. We propose a family of {\it semi-supervised estimators} \tcr{$\hvtss$ of $\vt$:}

%indexed by $\{\hphi(\cdot,\cdot),\hvti,\hf(\cdot)\}$.

\vspace{0.05in}
\tcb{In summary, we obtain our semi-supervised estimators $\hvtss$ of the quantile $\vt$ in three steps: }
\begin{enumerate}[(i)]
	\item \tcb{Calculate an initial estimator $\hvti$ of $\vt$ and an estimator $\hf(\cdot)$ of the density function $f(\cdot)$;}
	
	\item \tcb{Obtain the imputation function $\hphi(\X,\hvti)$ by the cross-fitting procedures \eqref{ds};}
	
	\item \tcb{Plug $\{\hf(\hvti),\hphi(\X,\hvti)\}$ in the one-step \tcmAC{update} formula \eqref{defi} \tcmAC{to obtain the final $\hvtss$}.}
\end{enumerate}

%\vspace{0.05in}
\begin{remark}[Robustifying and debiasing nature of the representation \eqref{defi}]\label{remark_robustifying_debiasing}
	\tcb{In addition to robustifying the estimator $\hvtss$, as discussed after \eqref{popee}, another effect of the term $\E_n\{\hphi(\X,\hvti) - \psi(Y,\hvti)\}$ in \eqref{defi} is eradicating the first-order error of $\hphi(\cdot,\cdot)$ as an estimator of $\phi(\cdot,\cdot)$ so that the influence function of $\hvtss$ is not affected by the estimation error of $\hphi(\cdot,\cdot)$. This property is crucial for ensuring the $n^{1/2}$-consistency and asymptotic normality of $\hvtss$, particularly when $\X$ is high dimensional or $\hphi(\cdot,\cdot)$ involves nonparametric calibrations. We will formally discuss this point in Remark \ref{remarkif} after obtaining the theoretical results of $\hvtss$. Interestingly, even if the imputation function satisfies $\E\{\phi(\X,\theta)\}=\E\{\psi(Y,\theta)\}$, this \emph{robustifying and debiasing} term should \emph{always} be included so that $\hvtss$ can enjoy the desirable properties mentioned above.}
\end{remark}

\tcb{\subsection{Main results: Theoretical properties of the semi-supervised estimators}\label{sec_main_results}}
The definition (\ref{defi}) now equips us with a family of semi-supervised estimators \tcr{$\hvtss$} for $\vt$ indexed by $\{\hphi(\cdot,\cdot),\hvti,\hf(\cdot)\}$. To study their limiting behavior, we assume the following conditions.

\vspace{0.05in}
\begin{assumption}
	\label{ainit}
	The estimators $\hvti$ and $\hf(\cdot)$ satisfy that
	\bse
	\hvti-\vt~=~O_p(u_{n}) \hbox{ and } \hf(\hvti)-f(\vt)~=~O_p(v_n)
	\ese
	for some positive sequences $u_n=o(1)$ and $v_n=o(1)$.
\end{assumption}

\begin{assumption}
	\label{aimp}
	%The second moment of $\phi(\X,\vt)$ is finite.
	\tcmAC{$\E[\{\phi(\X,\vt)\}^2] < \infty$.}
	%% NOTE: Changing this for two reasons: it's more clear, and it also helps with better spacing/formatting (along use of -ve vspace before Section 2.4) -- AC (1/16/2022).
	In addition, for any sequence $\tvt\to\vt$ in probability,
	\be
	\mbG_n\{\phi(\X,\tvt)-\phi(\X,\vt)\}~=~o_p(1) \tcm{ \hbox{ ~and~ }} \mbG_{n+N}\{\phi(\X,\tvt)-\phi(\X,\vt)\}~=~o_p(1).\label{uni1}
	\ee
\end{assumption}

\begin{assumption}
	\label{aest}
	Denote the estimation error
	$
	\hpsi(\X,\theta):=\hphik(\X,\theta)-\phi(\X,\theta)
	$
	and its second moment $\Delta_k(\cl):=(\sb\E_\X[\{\hpsi(\X,\theta)\}^2])^{1/2}\ (k=1,\ldots,\kK)$. Then\tcmAC{,} the set
	\be
	\mp_{n,k}~:=~\{\hpsi(\X,\theta):\theta\in\mbtv\} \tcr{,}
	\label{pnk}
	\ee
	\tcr{for some $\varepsilon > 0$,} satisfies that, for any $\eta\in(0,\Delta_k(\cl)+\xi\,]$ with some $\xi>0$,
	\be
	N_{[\,]}\{\eta,\mp_{n,k}\mid\cl,L_2(\P_\X)\}~\leq~ H(\cl) \eta^{-c} \quad (k=1,\ldots,\kK)\tcr{,}
	\label{vc}
	\ee
	with some function $H(\cl)>0$ such that $H(\cl)=O_p(a_n)$ for some positive sequence $a_n$, where the symbol $N_{[\,]}(\cdot,\cdot,\cdot)$ refers to the bracketing number defined in \citet{van1996weak} and \citet{van2000asymptotic}. Here $\mp_{n,k}$ is indexed by $\theta$ \tcr{{only}} and treats $\hpsi(\cdot,\theta)$ as a non\tcr{-}random function $(k=1,\ldots,\kK)$. Further, for some positive sequences $d_{n,2}$ and $d_{n,\infty}$ allowed to diverge,
	\be
	\Delta_k(\cl)~=~O_p(d_{n,2}) \tcm{\hbox{ ~and~ }} \sbx|\hpsi(\tcb{\x},\theta)|~=~O_p(d_{n,\infty}) \quad (k=1,\ldots,\kK). \label{psi_hat_norms}
	\ee
\end{assumption}

\begin{remark}\label{remarkass}
	Assumption \ref{ainit} is standard for one-step estimators, ensuring good behavior of $\hvti$ and $\hf(\cdot)$. Assumption \ref{aimp} outlines features of a reasonable imputation function $\phi(\cdot,\cdot)$. According to Example 19.7 and Lemma 19.24 of \citet{van2000asymptotic}, the condition (\ref{uni1})  is true provided $\phi(\X,\theta)$ is Lipschitz continuous in $\theta$. Assumption \ref{aest} is imposed to control the estimation error of $\hphik(\X,\theta)$ in the neighborhood $\mbtv$ of $\vt$. The condition (\ref{vc}) therein holds when $\mp_{n,k}$ is a VC class given $\cl$ \citep{van1996weak}. Also, we put the restriction (\ref{psi_hat_norms}) with possibly divergent rates on the $L_2$ and $L_\infty$ norms of $\hpsi(\X,\theta)$, \tcr{\emph{weaker}} than requiring its convergence uniformly over $\x\in\mx$ and $\theta\in\mb(\vt,\varepsilon)$, i.e., \tcm{the $L_\infty$ convergence}. All these (high-level) assumptions are fairly mild and will be verified for some choices of $\{\phi(\cdot,\cdot),\hphik(\cdot,\cdot)\}$ in Section \ref{secnf}; \tcr{see, e.g., Propositions \ref{thphi}--\ref{thbn} and the discussions after Theorem \ref{thhd} therein}.
	In addition, we do {\it not} assume $\lim_{n\to\infty}(n/N)= 0$, a common requirement in the semi-supervised literature \citep{chakrabortty2018efficient, gronsbell2018semia}. Our conclusions thus remain valid \tcr{even} when the labeled and unlabeled data are comparable in size. Nevertheless $\nu=0$ is a more practically relevant and theoretically challenging case, significantly different from the traditional missing data problem.
\end{remark}

In the following theorem, we state \tcr{the} large sample properties of $\hvtss$ defined by (\ref{defi}), \tcr{giving a complete characterization of its asymptotic expansion under general (high-level) conditions.}

\begin{theorem}[\tcr{General asymptotic expansion of $\hvtss$}]\label{thos}
	If Assumptions \ref{adensity}--\ref{aest} hold, then
	\bse
	\hvtss-\vt~=~\{nf(\vt)\}^{-1}\sl\omega_{n,N}(\Z_i,\vt)+O_p(u_n^2+u_nv_n+n^{-1/2}r_n)+o_p(n^{-1/2})\tcr{,}
	\ese
	where $\omega_{n,N}(\Z,\theta):= \phi(\X,\theta)-\psi(Y,\theta)-\E_{n+N}\{\phi(\X,\theta)\}$ satisfying $\E\{\omega_{n,N}(\Z,\theta)\}=0$, and
	\bse
	r_n:=d_{n,2}\{\log\,a_n+\log\,(d_{n,2}^{-1})\}+n_\kK^{-1/2}d_{n,\infty}\{(\log\,a_n)^2+(\log\,d_{n,2})^2\}.
	\ese
	Further, given
	\be
	u_nv_n+u_n^2+n^{-1/2}r_n~=~o(n^{-1/2}),
	\label{negligible}
	\ee
	the limiting distribution of $\hvtss$ is $n^{1/2}f(\vt)\sigss^{-1}(\hvtss-\vt)\to\mn(0,1)$  $(n,N\to\infty)$,
	where
	\bse
	\sigss^2~:=~\var\{\omega_{n,N}(\Z,\vt)\}~=~(1-\nu_{n,N})\var\{\psi(Y,\vt)-\phi(\X,\vt)\}+\nu_{n,N}\var\{\psi(Y,\vt)\}.
	\ese
\end{theorem}

\begin{remark}\label{rvarq}
	The asymptotic variance of $\hvtss$ can be estimated by $n^{-1}\{\hf(\hvti)\}^{-2}\hsigss^2$ with
	\be
	\hsigss^2~:=~(1-\nu_{n,N})\var_n\{\psi(Y,\hvti)-\hphi(\X,\hvti)\}+ \nu_{n,N}\var_n\{\psi(Y,\hvti)\}.
	\label{varq}
	\ee
	Of course, one can replace the initial estimator $\hvti$ in $\hf(\hvti)$ and (\ref{varq}) by the semi-supervised estimator $\hvtss$. The results %of
	\tcmAC{from our} simulations
	in Section \ref{secs}, however, show that $\{\hf(\hvti)\}^{-1}\hsigss$ works \tcmAC{quite}
	well for estimating $\{f(\vt)\}^{-1}
	\sigss$. In addition, using $\{\hf(\hvti)\}^{-1}\hsigss$ reduces %\tcmAC{the}
	computational burden, since $\hphi(\X_i,\hvti)$ is \tcmAC{already} available from \tcmAC{the}
	previous steps\tcmAC{,} while $\hphi(\X_i,\hvtss)$ $(i=1,\ldots,n)$ needs to be calculated \tcmAC{afresh} via the cross-fitting procedure \eqref{ds}.
\end{remark}

%\tcr{While Theorem \ref{thos} , as} the
As the most important special case of Theorem \ref{thos}, \tcm{which holds for {any} $\nu_{n,N}$ and its limit $\nu$ -- positive or zero},
the limiting behavior of $\hvtss$ when $\nu=0$ is considered in the next corollary.
\begin{corollary}[\tcr{Properties of $\hvtss$ under the special case $\nu = 0$}]\label{cor1}
	Assume that the conditions of Theorem \ref{thos} hold and that $\nu=0$\tcr{. Then,} the semi-supervised estimator $\hvtss$ \tcr{satisfies}\tcmAC{:} %is such that
	\bse
	&&\hvtss-\vt~=~\{nf(\vt)\}^{-1}\sl\omega(\Z_i,\vt)+O_p(u_n^2+u_nv_n+n^{-1/2}r_n)+o_p(n^{-1/2}),  \tcm{\hbox{ and }}\\
	&&n^{1/2}f(\vt)\tsigss^{-1}(\hvtss-\vt)~\to~\mn(0,1)\quad (n,N\to\infty),
	\ese
	where $\omega(\Z,\theta):= \phi(\X,\theta)-\psi(Y,\theta)-\E\{\phi(\X,\theta)\}$ satisfying $\E\{\omega(\Z,\theta)\}=0$, and
	\bse
	\tsigss^2~:=~\var[\{\omega(\Z,\vt)\}^2]~=~\var\{\psi(Y,\vt)-\phi(\X,\vt)\}.
	\ese
\end{corollary}

\begin{remark}[Robustness and first-order insensitivity]\label{remarkif}
	Theorem \ref{thos} presents the $n^{1/2}$-consistency and asymptotic normality of $\hvtss$ with an \emph{arbitrary} choice of $\{\phi(\cdot,\cdot),\hphik(\cdot,\cdot)\}$ under the assumptions therein. \tcr{In this sense,} %In semi-supervised settings,
	it provides a \tcr{\emph{family of globally robust semi-supervised estimators}} with influence functions indexed by $\phi(\cdot,\cdot)$. In addition, we observe that estimating $\phi(\cdot,\cdot)$ by $\hphi(\cdot,\cdot)$ does \tcr{\emph{not}} affect the influence function of $\hvtss$, as long as the high-level conditions in Assumption \ref{aest} are satisfied. Therefore, $\hvtss$ is first-order insensitive to estimation errors and \tcr{any knowledge of the construction} %construction knowledge
	of $\hphi(\cdot,\cdot)$. This property is particularly desirable for inference when $\X$ is high dimensional or $\hphi(\cdot,\cdot)$ involves nonparametric techniques %such as kernel smoothing
	\tcr{-- cases when it may not be $n^{-1/2}$-rate.}
\end{remark}

\begin{remark}[Efficiency \tcr{ comparisons, and some examples of the imputation function $\phi(\cdot,\cdot)$}]\label{remarkeff}
	If we take $\phi(\X,\theta)\equiv\E\{\psi(Y,\theta)\mid \bfd(\X)\}$ with some possibly unknown function $\bfd(\cdot)$, then
	\bse
	\sigss^2~\equiv~\E[\{\psi(Y,\vt)\}^2]-(1-\nu_{n,N})\E[\{\phi(\X,\vt)\}^2]~\leq~\sigsup^2,
	\ese
	i.e., the semi-supervised variance $\{f(\vt)\}^{-2}\sigss^2$ in Theorem \ref{thos} is no more than the supervised variance $\{f(\vt)\}^{-2}\sigsup^2$ in Proposition \ref{p1}, indicating $\hvtss$ is \tcr{\emph{equally or more efficient}} compared to the supervised estimator $\hvts$. An example of $\bfd(\x)$ is the linear transformation $\bfd(\x)\equiv\mbP_0\trans\x$, where $\mbP_0$ is some unknown $r\times p$ matrix, with a fixed $r\leq p$, \tcb{that can be chosen and estimated using parametric regression methods ($r=1$), e.g., linear regression of $Y$ vs. $\X$, or dimension reduction techniques ($r\geq \tcr{1}$) %($r\geq 2$),
		such as %the
		\tcr{sliced} inverse regression \citep{li1991sliced, lin2019sparse}; see Section \ref{secs} for the implementation details of estimating $\mbP_0$.} %{\color{magenta}\bf Do you want to point to somewhere later where this issue is taken up?}\tcb{**Done**}
	\tcb{After obtaining an estimator of $\mbP_0$, a further step of nonparametric smoothing can be conducted to approximate  $\phi(\x,\theta)\equiv\E\{\psi(Y,\theta)\mid \mbP_0\trans\X=\mbP_0\trans\x\}$.} \tcb{In Section \ref{secnf}, we will substantiate \tcr{that} %{\color{magenta}\bf where?}\tcb{**Done**}
		$\phi(\x,\theta)\equiv\E\{\psi(Y,\theta)\mid\mbP_0\trans\X=\mbP_0\trans\x\}$ and %some
		\tcr{its} corresponding nuisance
		estimators %\tcr{of such $\phi(\cdot)$}
		indeed satisfy the high-level conditions required in Theorem \ref{thos}.}

	Further, when $\phi(\X,\theta_0)=\E\{\psi(Y,\theta_0)\mid \X\}$ and $\nu>0$, we have\tcmAC{:}
	\be
	\sigss^{2}&~=~&(1-\nu_{n,N})\var[\psi(Y,\theta_0)-\E\{\psi(Y,\theta_0)\mid\X\}]+\nu_{n,N}\var\{\psi(Y,\vt)\} \nonumber\\
	&~\to~&(1-\nu)\E([\psi(Y,\theta_0)-\E\{\psi(Y,\theta_0)\mid\X\}]^2)+\nu\,\E[\{\psi(Y,\vt)\}^2]~=~\sigeff^2\tcr{,}\label{eff0}
	\ee
	with $\{f(\vt)\}^{-2}\sigeff^2$ the \tcr{\emph{semi\tcr{-}parametric efficiency bound}} for estimating $\vt$ in missing data theory \citep{tsiatis2007semiparametric,graham2011efficiency}. If $\phi(\X,\theta_0)=\E\{\psi(Y,\theta_0)\mid \X\}$ and $\nu=0$, Corollary \ref{cor1} gives\tcmAC{:}
	\bse
	\tsigss^2~=~\E([\psi(Y,\vt)-\E\{\psi(Y,\theta)\mid \X\}]^2)\leq \E[\{\psi(Y,\vt)-g(\X)\}^2]\tcr{,}
	\ese
	for any function $g(\cdot)$ \tcr{in $L_2(\P_{\X})$,}  %\tcr{with $\E\{g^2(\X)\} < \infty$,}
	and the equality holds only if $g(\X)=\E\{\psi(Y,\vt)\mid \X\}$ almost surely. This fact reveals the asymptotic \tcr{\emph{optimality}} of $\hvtss$ among all regular and asymptotically linear estimators of $\vt$, whose influence functions take the form\tcr{:} $\{f(\vt)\}^{-1}\{g(\X)-\psi(Y,\vt)\}$\tcr{,} for some function $g(\cdot)$. \tcr{Further, u}nder the \tcr{appropriate} semi\tcr{-}parametric model of $\Z$ \tcr{-- one} where the distribution of $\X$ is known while that of $Y$ is unrestricted up to Assumption \ref{adensity}, one can show \tcr{that} $\{f(\vt)\}^{-2}\E([\psi(Y,\vt)-\E\{\psi(Y,\vt)\mid \X\}]^2)$ equals the \emph{efficient} asymptotic variance for estimating $\vt$. %, %i.e.,
	\tcr{Thus, for \emph{any} $\nu \geq 0$,}
	%the estimator
	$\hvtss$ \tcr{\emph{achieves semi\tcr{-}parametric efficiency} with $\phi(\cdot,\cdot)$ as above.}
	%\tcr{(under this model)
		%when $\phi(\X,\theta_0)=\E\{\psi(Y,\theta_0)\mid \X\}$}.
\end{remark}

\section{Choice and estimation of the nuisance functions}\label{secnf}
This section details some choices and estimators of the imputation function $\phi(\cdot,\cdot)$ used in the construction of \tcr{our semi-supervised estimators} $\hvtss$ \tcr{in \eqref{defi}}. Although an arbitrary imputation function equips our method with the $n^{1/2}$-consistency and asymptotic normality stated in Theorem \ref{thos} if the high-level conditions in Assumption \ref{aimp} hold, the ideal choice from the perspective of efficiency is $\phi(\X,\theta)=\E\{\psi(Y,\theta)\mid\X\}$ as discussed in Remark \ref{remarkeff}. However, when the dimension $p$ of $\X$ is large, estimating the conditional mean $\E\{\psi(Y,\theta)\mid\X\}$ fully nonparametrically is generally undesirable \tcr{in practice due to curse of dimensionality that typically enforces}
%as it typically requires
stringent conditions such as undersmoothing \citep{chakrabortty2018efficient}. A common strategy is implementing \tcr{suitable} \tcr{\emph{dimension reduction}} techniques followed by \tcr{\emph{nonparametric calibrations}} targeting the function $\E\{\psi(Y,\theta)\mid \S\}$ rather than $\E\{\psi(Y,\theta)\mid\X\}$, where $\S:=\mbP_0\trans\X\in\ms\subset\rR^r$ and $\mbP_0$ is a $p\times r$ matrix with some fixed $r\leq p$. Here we emphasize that
$
\E\{\psi(Y,\theta)\mid\S\}=\E\{\psi(Y,\theta)\mid\X\}
$
is {\it not} assumed anyway, i.e., the dimension reduction is {\it not} necessarily sufficient. According to Remark \ref{remarkeff}, the advantage of $\hvtss$ over $\hvts$ in terms of efficiency is ensured by setting
\be
\phi(\X,\theta)~\equiv~\phi(\X,\theta,\mbP_0)~\equiv~\E\{\psi(Y,\theta)\mid \mbP_0\trans\X\}~\equiv~\E\{\psi(Y,\theta)\mid \S\}\tcr{,}
\label{phis}
\ee
regardless of whether the dimension reduction is sufficient or not. Hence $\mbP_0$ can be \tcr{\emph{any}} user-defined or data-dependent matrix. If $\mbP_0$ is entirely determined by $\P_\X$, we can \tcr{safely} assume its estimation error to be negligible due to the plentiful observations for $\X$ in semi-supervised settings. An example is the $r$ leading principal component directions of $\X$. However, to make the dimension reduction and the imputation as sufficient as possible, we mainly consider cases where $\mbP_0$ \emph{depends} on the joint distribution of $(Y,\X\trans)\trans$ and thereby needs to be estimated \tcr{on $\cl$} with significant errors. Some reasonable choices of such $\mbP_0$ will be discussed in Remark \ref{rkernel}.

\vspace{0.05in}
To justify the \tcr{usage of}
imputation function\tcr{s of the form} $\phi(\X,\theta)\equiv\E\{\psi(Y,\theta)\mid \S\}$ \tcr{as in \eqref{phis}},
we now show it satisfies Assumption \ref{aimp} under a mild condition.

\begin{proposition}\label{thphi}
	Suppose that $\E[\{\sb f(\theta\mid\S)\}^2]<\infty$ \tcr{for some $\varepsilon$,} with
	$f(\cdot\mid\S)$ the conditional density of $Y$ given $\S$. Then\tcm{,} Assumption \ref{aimp} is satisfied by $\phi(\X,\theta)\equiv\E\{\psi(Y,\theta)\mid\S\}$.
\end{proposition}

To approximate the conditional mean $\phi(\X,\theta)\equiv\E\{\psi(Y,\theta)\mid \mbP_0\trans\X\}$, we may employ any \tcr{suitable} smoothing technique, such as kernel smoothing, kernel machine regression and smoothing splines. For \tcr{sake of} illustration, we focus \tcr{here} on the \tcr{\emph{kernel smoothing estimator(s):}}
\be
&\hphik(\x,\theta)~\equiv~\hphik(\x,\theta,\hmbP)~:=~\{\hl(\x,\hmbP)\}^{-1}\hm(\x,\theta,\hmbP)\  (k=1,\ldots,\kK), \hbox{ where }
\label{ks} \\
&\tcm{\hl(\x,\mbP):=h_n^{-r}\Enk [K_h\{\mbP\trans(\x-\X)\}]\hbox{ and } \hm(\x,\theta,\mbP):=h_n^{-r} \Enk[ \psi(Y,\theta)K_h\{\mbP\trans(\x-\X)\}],} \nonumber
\ee
%where $\hl(\x,\mbP):=h_n^{-r}\Enk [K_h\{\mbP\trans(\x-\X)\}]$ and
%\bse
%\hm(\x,\theta,\mbP)~:=~h_n^{-r} \Enk[ \psi(Y,\theta)K_h\{\mbP\trans(\x-\X)\}]\tcr{,}
%\ese
with the notation $\E_{n,k}\{g(\Z)\}:=n_{\kK^-}^{-1}\slk g(\Z_i)$ for any function $g(\cdot)$, and with $\hmbP$ being {any} estimator of $\mbP_0$ based on the data set $\cl_k^-$, $K_h(\s):=K(h_n^{-1}\s) $, $K(\cdot):\rR^r\mapsto\rR$ a kernel function, e.g., \tcm {the standard Gaussian kernel,} and $ h_n\to 0$ denoting a bandwidth sequence. Considering $\X$ is possibly high dimensional and $\mbP_0$ needs to be estimated, establishing the \tcr{(uniform)} convergence \tcr{properties} %property
of $\hphik(\x,\theta,\hmbP)$ poses substantial technical challenges and has not been studied in the literature yet \tcr{to the best of our knowledge}. \tcb{In contrast, \tcr{most of} the existing works on kernel smoothing with \tcr{such estimated transformed covariates -- often termed as ``generated''}  covariates \tcm{-- e.g., \citet{mammen2012nonparametric}, \citet{escanciano2014uniform} and \citet{mammen_rothe_schienle_2016}}, mainly focus on the scenario where the dimension of $\X$ is fixed. \tcr{Our result below is thus \emph{novel} in this sense.}
	%In this sense, our result below is \emph{novel} and \emph{nontrivial}.}
}

\vspace{0.05in}
\tcb{To derive the convergence rate of $\hphik(\x,\theta,\hmbP)$, \tcr{as} %which is
defined in \eqref{ks}}, \tcr{\emph{uniformly}} over $\x\in\mx$ and $\theta\in\mbtv$, we impose the following requirements, along with some standard smoothness conditions for kernel smoothing \tcr{which are listed} in Assumption \ref{akernel} of Appendix \ref{sm_assumptions}.

\begin{assumption}\label{al1}
The estimation error $\|\hmbP-\mbP_0\|_1=O_p(\alpha_n)$ for some positive sequence $\alpha_n$.
\end{assumption}

%\begin{assumption}\label{akernel} (smoothness conditions)
%(i) The function $K(\cdot):\rR^r\mapsto\rR$ is a symmetric kernel of order $d\geq 2$ with a finite $d$th moment. Moreover, it  is bounded, integrable and continuously differentiable. In addition, denote $\nabla K(\s):=\partial K(\s)/\partial \s$. Then there exists some constant $v> 1$ such that $\|\nabla K(\s)\|\leq c_1\,\|\s\|^{-v}$ for any $\|\s\|>c_2$. (ii) The support $\ms$ of $\S\equiv\mbP_0\trans\X$ is compact. The density function $f_{\S}(\cdot)$ of $\S$ is bounded and bounded away from zero on $\ms$. In addition, it is $d$ times continuously differentiable with a bounded $d$th derivative on some open set $\ms_0\supset\ms$. (iii) With respect to $\s$, the conditional distribution function $F(\theta\mid\S=\s)$ of $Y$ given $\S=\s$ is $d$ times continuously differentiable and has a bounded $d$th derivatives on $\ms_0\times\mbtv$.
%\end{assumption}

\begin{assumption}[\tcr{Required only when $\mbP_0$ needs to be estimated}]\label{ahbe} (i) The support $\mx$ of $\X$ is such that $\sx\|\x\|_\infty<\infty$. (ii)  The function $\nabla K(\s):=\partial K(\s)/\partial\s$ is continuously differentiable and satisfies $\|\partial \{\nabla K(\s)\}/\partial \s\|\leq c\,\|\s\|^{-w}$ for any $\|\s\|>c$, where $w> 1$ is some constant. Further, it is locally lipschitz continuous, i.e., $\|\nabla K(\s_1)-\nabla K(\s_2)\|\leq \|\s_1-\s_2\|\rho(\s_2)$ for any $\|\s_1-\s_2\|\leq c$, where $\rho(\cdot)$ is some bounded and square integrable function with a bounded derivative $\nabla\rho(\cdot)$. (iii) Let $\bfeta_{t[j]}(\s,\theta)$ be the $j$th component of $\bfeta_{t}(\s,\theta):=\E[\X\{\psi(Y,\theta)\}^t\mid \S=\s]$. Then, with respect to $\s$, the function $\bfeta_{t[j]}(\s,\theta)$ is continuously differentiable and has a bounded first derivative on $\ms_0\times\mbtv$ $(t=0,1;\,j=1,\ldots p)$ for some open set $\ms_0\supset\ms$.
\end{assumption}

%\rjccomment{There are a lot of {\it In the above} phrases. The term needs to be more precise as to where. \tcb{Now there is only one ``in the above'' in the whole draft.} Now there are none!!}

Assumption \ref{al1} regulates the behavior of $\hmbP$ as an estimator of the transformation matrix $\mbP_0$. Assumption \ref{ahbe} requires mild smoothness conditions to control the estimation error of $\hmbP$ while (ii) \tcr{therein} is satisfied \tcr{in particular} by the second-order Gaussian kernel, among others. Similar assumptions can be found in \citet{chakrabortty2018efficient} that studied kernel smoothing estimators with dimension reduction in low dimensional scenarios. We now propose the following result.

\begin{theorem}[\tcr{Convergence of $\hphik$}]\label{thhd}
Set $\gamma_n:=[(nh_n^r)^{-1}\max\{\log(h_n^{-r}),\log(\log\,n)\}]^{1/2}$, $s_{n,1}:=\gamma_n+h_n^d$ and $s_{n,2}:=h_n^{-2}\alpha_n^2+h_n^{-1}\gamma_n\alpha_n+\alpha_n$. If Assumptions \ref{al1}--\ref{ahbe}, as well as Assumption \ref{akernel} in Appendix \ref{sm_assumptions}, hold and $s_{n,1}+s_{n,2}=o(1)$, then
\bse
\sbx|\hphik(\x,\theta,\hmbP)-\phi(\x,\theta,\mbP_0)| ~=~O_p(s_{n,1}+s_{n,2}) \quad (k=1,\ldots,\kK).
\ese
\end{theorem}

\begin{remark}[%\tcb{Uniform convergence of $\hphik(\x,\theta,\hmbP)$}
\tcr{Convergence rates -- examples of $\hmbP$}]\label{rkernel}
Theorem \ref{thhd} establishes the $L_\infty$ error rate of $\hphik(\x,\theta,\hmbP)$ under mild conditions. The uniform consistency of $\hphik(\x,\theta,\hmbP)$ is ensured at the optimal bandwidth rate $\hopt=O\{n^{-1/(2d+r)}\}$ for any kernel order $d\geq 2$ and fixed $r$, if
\be
\alpha_n~=~o\{n^{-1/(2d+r)}\}.
\label{alpharate}
\ee
We consider the validity of \eqref{alpharate} for \tcr{some} common choices of $\mbP_0$, such as the least square\tcr{s} regression parameter $(r=1)$ satisfying
$
\E\{\X(Y-\mbP_0\trans\X)\}=\bze_p,
$
%and
\tcr{or} the $r$ leading eigenvectors \tcr{($r \geq 1$)} of the matrix $\cov\{\E(\X\mid Y)\}$, which can be estimated by %the
\tcr{sliced} inverse regression \citep{li1991sliced}. When $p$ is fixed, there typically exist $n^{1/2}$-consistent estimators $\hmbP$ of $\mbP_0$\tcr{, so} \eqref{alpharate} is satisfied by the fact that $\alpha_n=O(n^{-1/2})$. In high dimensional scenarios where $p$ is divergent and greater than $n$, one can obtain $\hmbP$ \tcr{based on} %from
%the
regularized version\tcr{s} of linear regression or %the
\tcr{sliced} inverse regression \citep{lin2019sparse}. \tcr{In these cases, the} %The
sequence $\alpha_n=O\{q(\log\,p/n)^{1/2}\}$ when the \tcb{$L_1$ penalty} is used under some suitable conditions \tcr{\citep{buhlmann2011statistics,wainwright2019high, lin2019sparse}}, %negahban2012unified,
where $q:=\|\mbP_0\|_0$ represents the sparsity level. Hence (\ref{alpharate}) is true when\tcr{ever} $q(\log\,p)^{1/2}=o\{n^{(2d+r-2)/(4d+2r)}\}$. \tcb{An example of $\hmbP$, with $r=1$, for high dimensional data is the lasso estimator} $\tcb{\hmbP\equiv\hbox{$\arg\min_{\bbeta\in\rR^p}$}\{n_{\kK^-}^{-1}\hbox{$\sum_{i\in\I_k^-}$}(Y_i-\bbeta\trans\X_i)^2+\lambda_{n,k}\|\bbeta\|_1\}}$, \tcb{where $\lambda_{n,k}>0$ is a tuning parameter. After this parametric regression step, we conduct kernel smoothing of $\psi(Y,\hvti)$ on the one-dimensional linear combination $\hmbP\trans\X$ to obtain $\hphik(\x,\hvti,\hmbP)$, as in \eqref{ks}; see Section \ref{secs} for the details of constructing such a high-dimensional nuisance estimator.}
\end{remark}

Theorem \ref{thhd} implies that $d_{n,2}=o(1)$ and $d_{n,\infty}=o(1)$ for Assumption \ref{aest} when $\phi(\x,\theta,\mbP)$ and $\hphik(\x,\theta,\mbP)$ are as in (\ref{phis})--(\ref{ks}). We now validate the condition (\ref{vc}) on the bracketing number.
\begin{proposition}\label{thbn}
Suppose that $\E[\{\sb f(\theta\mid\S)\}^2]<\infty$\tcmAC{,} \tcr{for some $\varepsilon > 0$}. Then\tcmAC{,} the set $\mp_{n,k}$ defined in \eqref{pnk} satisfies
$
N_{[\,]}\{\eta,\mp_{n,k}\mid\cl,L_2(\P_\X)\}\leq c\,(n+1)\eta^{-1}.
$
\end{proposition}

\begin{remark}[\tcb{Verification of the condition \eqref{negligible}}]\label{rcondition}
The results of Theorem \ref{thhd} and Proposition \ref{thbn} indicate \tcr{that} $a_n=O(n)$, $\ d_{n,2}=o(1)$ and $\ d_{n,\infty}=o(1)$ in Assumption \ref{aest}. Moreover, by setting the initial estimator $\hvti\equiv\hvts$ and estimating the density function $f(\cdot)$ via kernel smoothing with a second-order kernel function at the optimal bandwidth rate, we have $u_n=O(n^{-1/2})$ and $\ v_n=O\{(n^{-1}\log\,n)^{2/5}\}$ from Proposition \ref{p1} and \citet{hansen2008uniform}, where $\{u_n, v_n\}$ are as defined in Assumption \ref{ainit}. These results on the convergence rates of $\{a_n, d_{n,2}, d_{n,\infty}, u_n,v_n\}$ are actually sufficient for the condition (\ref{negligible}) in Theorem \ref{thos} and thus ensure the asymptotic normality of $\hvtss$.
\end{remark}

\begin{remark}[\tcb{Other reasonable \tcr{choices of} the nuisance estimator}]\label{remark_other_nuisance_functions}
	\tcmAC{As we conclude this section, we would like to reiterate that a}lthough we have focused on the combination of kernel smoothing and dimension reduction in this section, it is just {one} suitable strategy for approximating the nuisance function $\phi(\X,\theta)$ in \eqref{phis}. A wide class of alternatives could also be leveraged to estimate $\phi(\X,\theta)$ as long as the high-level conditions in Theorem \ref{thos} are satisfied. For instance, we can let $\mbP_0$ equal the $p\times p$ identity matrix in \eqref{phis} and approximate $\phi(\X,\theta)\equiv\E\{\psi(Y,\theta)\mid\X\}$ by popular machine learning approaches, such as random forest \citep{breiman2001random} and kernel machine regression \citep{liu2007semiparametric}, without use of dimension reduction. We will present the implementation details and numerical results related to random forest in Sections \ref{secs} and \ref{secda} while not \tcr{delving} %digging
	any further into the theoretical aspects, which are beyond the main interest of this article.
\end{remark}

\section{Simulations}\label{secs}
We study in this section the numerical performance of our \tcr{proposed semi-supervised inference} method on simulated data \tcr{and compare it  to the supervised counterpart}. Throughout, the quantile level is $\tau=0.5$. \tcb{The results are similar for other quantile levels, such as $\tau=0.25$ or $0.75$. We do not present them here for brevity.}
%{\color{magenta}\bf You might want to say the results are similar for $\tau = 0.25, 0.75$.}\tcb{**Done**}
We set the sample sizes $n=200$, $500$ or $2,000$ and $N=5,000$. The covariates $\X$ are  drawn from a $p$-dimensional normal distribution with a zero mean and an identity covariance matrix, where \tcr{we choose} $p=10$, $20$, $200$ or $500$. The conditional outcome model is chosen as $Y\mid\X\sim\mn\{m(\X),1\}$ for a variety of $m(\cdot)$ discussed below.
%{\color{magenta}\bf Maybe make $\mn$ to be $\hbox{Normal}$} \tcb{**It is $\hbox{Normal}$ now.**}

Let $\X_q:=(\X_{[1]},\ldots,\X_{[q]})\trans$, where $q=p$ when $p\in\{10, 20\}$, and $q=5$ or $\ceil{p^{1/2}}$ when $p\in\{200, 500\}$, %representing
\tcr{represents the} (effective) sparsity (fully dense for $p\in\{10,20\}$, and sparse or moderately dense for $p\in\{200,500\}$, respectively) of the true conditional mean model \tcr{$m(\X)$}, %given by\tcr{:}
\tcr{which we \tcm{set} as:} % choose as % DIfferent from the journal version. -- Dai 01/24/22
\begin{enumerate}[(a)]
	\item $m(\X)~\equiv0~$, {\it a null model};

	\item $m(\X)~\equiv~\bon_q\trans\X_q$, {\it a linear model};
	
	\item $m(\X)~\equiv~\bon_q\trans\X_q+(\bon_q\trans\X_q)^2/q$, {\it a single index model};
	
	\item $m(\X)~\equiv~(\bon_q\trans\X_q)\{1+2(\bze_{q-\ceil{q/2}}\trans,\bon_{\ceil{q/2}}\trans)\X_q/q\}$, {\it a double index model};
	
	\item $m(\X)~\equiv~\bon_q\trans\X_q+\|\X_q\|^2/3$, {\it a quadratic model}.
\end{enumerate}
%(a) $0$, {\it a null model}; (b) $\bon_q\trans\X_q$, {\it a linear model}; (c) $\bon_q\trans\X_q+(\bon_q\trans\X_q)^2/q$, {\it a single index model}; (d) $(\bon_q\trans\X_q)\{1+2(\bze_{q-\ceil{q/2}}\trans,\bon_{\ceil{q/2}}\trans)\X_q/q\}$, {\it a double index model}; \tcr{and} (e) $\bon_q\trans\X_q+\|\X_q\|^2/3$, {\it a quadratic model}.
These models generally represent a broad class of relation between $Y$ and $\X$, containing commonly encountered linear and non\tcr{-}linear (quadratic and interaction) effects, in both low and high dimensional scenarios. In each configuration, estimation and inference results of our semi-supervised estimators are summarized from 500 replications. In the interest of space, \tcr{the results} %outcomes
for $p=10$ or $200$ are given in Appendix \ref{sm_additional_simulations}.

For any kernel smoothing steps involved, we always use the second-order Gaussian kernel and \tcb{select the bandwidths} %{\color{magenta}\bf what likelihood?}\tcb{**Done**}
\tcb{by maximizing the \tcr{following cross-validated} likelihood function\tcr{:}
	\bse
	L(b)~:=~\hbox{$\prod_{i\in\I_k^-}$}\{\hphik^{(-i,b)}(\X_i,\hvti)+\tau\}^{I(Y_i<\hvti)}[1-\{\hphik^{(-i,b)}(\X_i,\hvti)+\tau\}]^{I(Y_i\geq\hvti)},
	\ese
	where $\hphik^{(-i,b)}(\cdot,\cdot)$ is a leave-one-out version of \eqref{ks} constructed with the data $\{(Y_j,\X_j\trans)\trans:j\in\I_k^-\backslash\{i\}\}$ $(k=1,\ldots,\kK)$ and the bandwidth $h_n\equiv b$.} All regularized approaches are based on the \tcb{$L_1$} \tcb{penalty} with tuning parameters chosen by ten-fold cross validation. % The number of folds in the cross-fitting process \eqref{ds} \tcr{as} is set as $\kK=10$. % DIfferent from the journal version. -- Dai 01/24/22
    \tcm{We set the number of folds in}

\begin{table}[H]
	\def~{\hphantom{0}}
	\caption{\tcb{Simulation results of Section \ref{secs}:} Efficiencies of \tcr{the} semi-supervised estimators relative to the supervised estimator. The \textbf{boldface} in each case represents the best efficiency\tcr{.}}
	\vskip 2mm
	
	{
		\resizebox{\textwidth}{!}{
			\begin{tabular}{c|c|ccccc||cccc||cccc}
				\hline
				\hline
				&         & \multicolumn{5}{c||}{$p=20$}           & \multicolumn{4}{c||}{$p=500$, $q=5$} & \multicolumn{4}{c}{$p=500$,   $q=\ceil{p^{1/2}}$} \\
				$n$                   & $m(\X)$ & KS$_1$ & KS$_2$ & PR   & RF   & ORE  & KS$_1$   & KS$_2$  & PR    & ORE   & KS$_1$      & KS$_2$      & PR        & ORE       \\ \hline
				\multirow{5}{*}{200}  & (a) & 0.90          & 0.93          & 0.86          & \textbf{0.94} & 1.00 & 0.95          & 0.68          & \textbf{0.99} & 1.00 & 0.95          & 0.68          & \textbf{0.99} & 1.00 \\
				& (b) & \textbf{4.76} & 4.36          & 2.58          & 1.38          & 4.38 & \textbf{2.31} & 1.41          & 1.88          & 2.50 & \textbf{2.96} & 1.14          & 1.22          & 4.62 \\
				& (c) & \textbf{4.23} & 4.03          & 2.74          & 1.37          & 4.36  & \textbf{1.78} & 1.02          & 1.44          & 2.21 & \textbf{1.95} & 1.07          & 1.24          & 4.60 \\
				& (d) & \textbf{3.71} & 3.40          & 2.48          & 1.34          & 4.11 & \textbf{1.37} & 0.85          & 1.34          & 2.12 & \textbf{1.77} & 1.05          & 1.24          & 4.24 \\
				& (e) & \textbf{2.45} & 2.31          & 1.99          & 1.38          & 4.74 & \textbf{1.90} & 1.24          & 1.56          & 2.65 & \textbf{1.20} & 0.91          & 1.17          & 4.97 \\
				%&     &               &               &               &               &      &  &               &               &               &      &  &               &               &               &      \\
				\hline
				\multirow{5}{*}{500}  & (a) & 0.98          & 0.95          & \textbf{0.98} & 0.98          & 1.00 & 0.96          & 0.91          & \textbf{0.99} & 1.00 & 0.96          & 0.91          & \textbf{0.99} & 1.00 \\
				& (b) & \textbf{3.99} & 3.78          & 3.55          & 1.45          & 3.70 & \textbf{2.64} & 2.53          & 2.39          & 2.31 & \textbf{3.90} & 3.22          & 2.32          & 3.86 \\
				& (c) & 3.76          & \textbf{3.96} & 3.54          & 1.43          & 3.69 & \textbf{2.13} & 1.94          & 1.68          & 2.07 & \textbf{3.56} & 3.05          & 2.35          & 3.84 \\
				& (d) & \textbf{3.48} & 3.31          & 3.36          & 1.44          & 3.51 & 1.65          & \textbf{1.71} & 1.56          & 2.00 & \textbf{3.21} & 2.89          & 2.25          & 3.60 \\
				& (e) & \textbf{2.38} & 2.32          & 2.24          & 1.47          & 3.93 & \textbf{1.70} & 1.63          & 1.68          & 2.43 & \textbf{2.00} & 1.89          & 1.58          & 4.09 \\
				%&     &               &               &               &               &      &  &               &               &               &      &  &               &               &               &      \\
				\hline
				\multirow{5}{*}{2000} & (a) & \textbf{1.00} & 1.00          & 0.99          & 0.99          & 1.00 & \textbf{1.00} & 0.96          & 1.00          & 1.00 & \textbf{1.00} & 0.96          & 1.00          & 1.00 \\
				& (b) & 2.65          & \textbf{2.66} & 2.61          & 1.45          & 2.34 & \textbf{1.76} & 1.75          & 1.74          & 1.80 & \textbf{2.29} & 2.28          & 2.04          & 2.39 \\
				& (c) & 2.66          & \textbf{2.69} & 2.65          & 1.45          & 2.34 & 1.57          & \textbf{1.57} & 1.47          & 1.69 & 2.23          & \textbf{2.29} & 2.04          & 2.39 \\
				& (d) & 2.59          & \textbf{2.62} & 2.58          & 1.46          & 2.28 & 1.48          & \textbf{1.56} & 1.42          & 1.65 & 2.15          & \textbf{2.16} & 1.98          & 2.31 \\
				& (e) & \textbf{1.72} & 1.71          & 1.69          & 1.38          & 2.42 & \textbf{1.65} & 1.64          & 1.63          & 1.86 & 1.72          & \textbf{1.73} & 1.63          & 2.46 \\
				\hline
	\end{tabular}}}
	\label{table_simulation_efficiency}
	
	\begin{footnotesize}
		\tcr{\underline{Glossary of notation}:} $p$, the dimension of $\X$; $q$, the sparsity level; $n$, the labeled data size; $m(\X)\equiv\E(Y\mid\X)$; KS$_1$/KS$_2$, kernel smoothing on the one/two direction(s) selected by linear regression/sliced inverse regression; RF, random forest; PR, parametric regression; ORE, oracle relative efficiency.
	\end{footnotesize}
\end{table}
\vspace{-0.05in}

\noindent \tcm{the cross-fitting process \eqref{ds} \tcr{as} $\kK=10$}. %Based on the labeled data $\cl$, the initial estimator $\hvti$ equals the sample median while $\hf(\cdot)$ is a kernel density estimator.
%\tcr{Using} the labeled data $\cl$, the
\tcr{T}he initial estimator $\hvti$  \tcr{in \eqref{defi} is chosen as} the sample median while $\hf(\cdot)$ is \tcr{taken as the} kernel density estimator of $Y$, \tcr{both obtained using $\cl$}.

To approximate the nuisance function $\phi(\x,\vt)$, the estimator $\hphik(\x,\hvti)$ leveraging the data $\cl_k^-$ is calculated by:
\vspace{-0.05in}
\begin{enumerate}[(I)]
	\item {\it kernel smoothing} \tcr{as} in \eqref{ks}, where the $p\times r$ transformation matrix $\hmbP$ is chosen as\tcr{:}
	\begin{enumerate}[(i)]
		\item the slope vector from unregularized or regularized linear regression ($r=1$) of $Y$ vs. $\X$,
		
		\item or the first two directions selected by the unregularized (with $\ceil{n/5}$ slices of equal width) or regularized (with $\ceil{n/75}$ slices of equal size) \tcr{versions of} sliced inverse regression ($r=2$) \citep{li1991sliced, lin2019sparse} of $Y$ vs. $\X$;
	\end{enumerate}

\item {\it parametric regression}, giving $\hphik(\x,\hvti)\equiv[1+\exp\{-(1,\x\trans)\hga_{k}\}]^{-1}-\tau$ with $\hga_{k}$ \tcr{being} the slope vector from unregularized or regularized logistic regression of $I(Y<\hvti)$ vs. $\X$;

\item {\it random forest} (for $p=10$ or $20$ only), treating $\psi(Y,\hvti)$ as the response,  growing $500$ trees and randomly sampling $\ceil{p^{1/2}}$ covariates as candidates at each split.
\end{enumerate}
Here\tcmAC{,} regularization is applied when $p\in\{200,500\}$. With these choices of $\hphik(\X,\hvti)$, incorporating a variety of flexible and easy-to-implement (parametric, semi\tcr{-}parametric or nonparametric) approaches to fitting (working) models between a continuous or binary response and a set of possibly high dimensional covariates, we construct the nuisance estimators $\hphi(\X,\hvti)$ via the cross-fitting process \eqref{ds}. Then our semi-supervised estimators $\hvtss$ are obtained by plugging $\{\hf(\hvti),\hphi(\X,\hvti)\}$ in the one-step formula \eqref{defi}.

\begin{comment}
\begin{enumerate}[(I)]
\item Kernel smoothing \eqref{ks}, where the $p\times r$ matrix $\hmbP$ is (i) the slope vector from unregularized or regularized linear regression ($r=1$) of $Y$ vs. $\X$; (ii) the first two directions selected by the unregularized (with $\ceil{n/5}$ slices of equal width) or regularized (with $\ceil{n/75}$ slices of equal size) sliced inverse regression ($r=2$) \citep{li1991sliced, lin2019sparse} of $Y$ vs. $\X$.

\item Parametric regression, giving $\hphik(\x,\hvti)\equiv[1+\exp\{-(1,\x\trans)\hga_{k}\}]^{-1}-\tau$ with $\hga_{k}$ the slope vector from unregularized or regularized logistic regression of $I(Y<\hvti)$ vs. $\X$.

\item Random forest (for $p=10$ or $20$ only), treating $\psi(Y,\hvti)$ as the response,  growing $500$ trees and randomly sampling $\ceil{p^{1/2}}$ covariates as candidates at each split.
\end{enumerate}
\end{comment}

\begin{table}[H]
	\def~{\hphantom{0}}
	\caption{\tcb{Simulation results of Section \ref{secs}:} Inference based on  the semi-supervised estimators using kernel smoothing on the direction selected by linear regression. \underline{All the numbers have been multi-} \underline{plied by 100}. The \textbf{boldfaces} are the coverage rates of \tcr{the} 95\% confidence intervals\tcr{.}}
	\vskip 2mm
	
	{
		\resizebox{\textwidth}{!}{
			\begin{tabular}{c|c|cccc||cccc||cccc}
				\hline
				\hline
				&         & \multicolumn{4}{c||}{$p=20$}& \multicolumn{4}{c||}{$p=500$, $q=5$}& \multicolumn{4}{c}{$p=500$,   $q=\ceil{p^{1/2}}$} \\
				$n$                   & $m(\X)$ & ESE   & Bias & ASE  & CR  & ESE     & Bias   & ASE    & CR    & ESE        & Bias       & ASE        & CR         \\
				\hline
				\multirow{5}{*}{200}  & (a) & 9.3  & 1.2  & 9.8  & \textbf{96.6}& 9.1  & -0.7 & 9.6  & \textbf{96.6}& 9.1  & -0.7 & 9.6  & \textbf{96.6} \\
				& (b) & 19.0 & -0.7 & 21.7 & \textbf{97.2}& 14.2 & 1.0  & 15.2 & \textbf{95.4}& 24.0 & 2.6  & 25.7 & \textbf{97.0} \\
				& (c) & 19.9 & -1.8 & 21.4 & \textbf{95.8}& 14.4 & 0.8  & 15.4 & \textbf{95.4}& 29.7 & 1.6  & 29.9 & \textbf{94.0} \\
				& (d) & 20.2 & 4.1  & 22.0 & \textbf{97.2}& 15.6 & 0.7  & 18.8 & \textbf{98.2}& 28.1 & 1.2  & 30.5 & \textbf{96.6} \\
				& (e) & 30.0 & -4.3 & 32.6 & \textbf{95.2}& 17.4 & -0.3 & 18.3 & \textbf{96.0}& 41.5 & 3.3  & 43.1 & \textbf{96.0} \\
				\hline
				%&     &      &      &      & \textbf{}    &      &      &      & \textbf{}    &      &      &      & \textbf{}     \\
				\multirow{5}{*}{500}  & (a) & 5.8  & 0.7  & 6.0  & \textbf{94.0}& 5.4  & 0.1  & 5.9  & \textbf{96.6}& 5.4  & 0.1  & 5.9  & \textbf{96.6} \\
				& (b) & 12.8 & -1.0 & 14.2 & \textbf{97.2}& 8.7  & -0.1 & 9.6  & \textbf{97.0}& 14.0 & 2.1  & 15.1 & \textbf{95.8} \\
				& (c) & 13.1 & -1.5 & 13.8 & \textbf{95.8}& 8.3  & 0.7  & 9.0  & \textbf{96.4}& 14.6 & 1.7  & 15.4 & \textbf{95.2} \\
				& (d) & 12.8 & 1.6  & 13.8 & \textbf{96.8}& 9.5  & 0.4  & 11.0 & \textbf{98.0}& 14.0 & 2.2  & 15.8 & \textbf{96.6} \\
				& (e) & 18.5 & -0.3 & 20.5 & \textbf{95.6}& 11.0 & -0.3 & 11.2 & \textbf{95.6}& 21.7 & 1.5  & 22.6 & \textbf{97.0} \\
				\hline
				%&     &      &      &      & \textbf{}    &      &      &      & \textbf{}    &      &      &      & \textbf{}     \\
				\multirow{5}{*}{2000} & (a) & 2.8  & 0.4  & 2.9  & \textbf{96.2}& 2.8  & -0.1 & 2.9  & \textbf{95.0}& 2.8  & -0.1 & 2.9  & \textbf{95.0} \\
				& (b) & 8.4  & -1.2 & 8.6  & \textbf{95.0}& 5.5  & 0.0  & 5.3  & \textbf{94.4}& 9.4  & 1.5  & 9.2  & \textbf{94.8} \\
				& (c) & 8.2  & -1.6 & 8.4  & \textbf{95.4}& 4.9  & 0.6  & 4.7  & \textbf{95.0}& 9.3  & 1.5  & 9.0  & \textbf{94.4} \\
				& (d) & 7.8  & 0.6  & 8.1  & \textbf{96.2}& 5.1  & 0.0  & 5.5  & \textbf{95.2}& 8.6  & 2.0  & 8.9  & \textbf{95.8} \\
				& (e) & 10.7 & -1.8 & 11.0 & \textbf{95.6}& 5.9  & -0.3 & 6.0  & \textbf{94.8}& 12.1 & 1.4  & 11.9 & \textbf{95.0} \\
				\hline
	\end{tabular}}}
	\label{table_simulation_inference}
	
	\begin{footnotesize}
		\tcr{\underline{Glossary of notation}:} $p$, the dimension of $\X$; $q$, the sparsity level; $n$, the labeled data size; $m(\X)\equiv\E(Y\mid\X)$; ESE, empirical standard error; ASE, average of estimated standard errors; CR, coverage rate of 95\% confidence intervals.
	\end{footnotesize}
\end{table}
\vspace{-0.05in}

%\noindent
Table \ref{table_simulation_efficiency}, along with Table \ref{sm_table_simulation_efficiency} of Appendix \ref{sm_additional_simulations}, presents the relative efficiencies, \tcm{given by:}
\bse
\E\{(\hvtss-\vt)^2\}/\E\{(\hvts-\vt)^2\},
\ese
of our semi-supervised estimators to the supervised estimator, i.e., the sample median of the labeled data $\cl$. For reference, we also provide the oracle relative efficiency $\sigsup^2/\sigeff^2$ given by Proposition \ref{p1} and \eqref{eff0}, which is achievable only asymptotically when the imputation function $\phi(\X,\theta)=\E\{\psi(Y,\theta)\mid\X\}$. The true values of $\sigsup$, $\sigeff$ and $\vt$ are approximated by Monte Carlo based on $100,000$ observations for $(Y,\X\trans)\trans$ independent of $\cl\cup\cu$. Except for the null model (a), where the unlabeled data $\cu$ does not help estimate $\vt$ in theory, the various estimators based on kernel smoothing or random forest generally outperform the supervised method across all scenarios. These results coincide with the discussion \tcr{on} %of
efficiency in Remark \ref{remarkeff} considering that both kernel smoothing and random forest target the function $\E\{\psi(Y,\vt)\mid\mbP_0\trans\X\}$ for some matrix $\mbP_0$. Also, the parametric regression approach with imputation functions from logistic regression shows superiority over the sample median of $\cl$ under all the models other than (a), indicating that our approach is fully robust and that the logistic model captures a part of the relation between $Y$ and $\X$. Moreover, we notice that the relative efficiencies of our estimators become closer to the corresponding oracle quantities as $n$ increases, verifying the \tcm{asymptotic} optimality claimed in Remark \ref{remarkeff}. In summary, these observations demonstrate the efficiency gain achieved by our semi-supervised method relative to its supervised competitor.

Next, in Table \ref{table_simulation_inference} as well as Table \ref{sm_table_simulation_inference} of Appendix \ref{sm_additional_simulations}, we display, as a representative case, the \tcr{results} %outcomes
of inference concerning $\vt$ based on our semi-supervised estimator \tcr{$\hvtss$ with $\hphik(\cdot)$ constructed} using kernel smoothing on the direction selected by linear regression. We report the bias, the empirical standard error, the average of the estimated standard errors, and the coverage rate of the 95\% confidence intervals, \tcm{based on the asymptotic normality from Theorem \ref{thos}}. We can see %{\color{magenta}\bf can you see it, or not?}\tcb{**See below for answer**}
that the biases are negligible, that the averages of the estimated standard errors are close to the corresponding empirical standard errors, and that the coverage rates are all around the nominal level \tcr{of} $0.95$.
%while the various choices of the function $\hphik(\X,\theta)$ highlight the full robustness and $n^{1/2}$-consistency of our approach.
Surprisingly, when the sample size $n=200$, the dimension $p=500$ and the sparsity level $q=\ceil{p^{1/2}}>n^{1/2}$, our method still generates satisfactory results and therefore shows insensitivity to the condition $\alpha_n=o(1)$ required by Theorem \ref{thhd}, considering %\tcr{that}
$\alpha_n=q(\log p/n)^{1/2}$ in Assumption \ref{al1} when the \tcb{$L_1$ penalty} is leveraged under some suitable conditions \citep{buhlmann2011statistics, negahban2012unified, wainwright2019high}. %{\color{magenta} The next is a very sweeping assertion \tcb{**Changed now**}}
Generally speaking, \tcb{the numbers in these tables, which are yielded by the inference procedures based on the limiting distribution in Theorem \ref{thos} and the variance estimate in Remark \ref{rvarq}, validate the theoretical results obtained in Section \ref{secos}.}
%justify the inference procedures for $\vt$ based on the limiting distribution in Theorem \ref{thos} and the variance estimate in Remark \ref{rvarq}.
With other choices of $\hphik(\cdot,\cdot)$, our method yields inference results similar in flavor to those in Tables \ref{table_simulation_inference} and \ref{sm_table_simulation_inference}. We thus omit them for the sake of brevity.

%\daicomment{Answer to ``can you see it, or not?'': The numbers in Table \ref{table_simulation_inference} have been multiplied by 100, so the biases are actually very close to zero and the difference between ESE and ASE is quite small in each row.}

\section{Real Data Analysis}\label{secda}

In this section, we apply our semi-supervised method to a subset of data from the National Health and Nutrition Examination Survey Data I Epidemiologic Follow-up Study, a study jointly initiated by the National Center for Health Statistics and the National Institute on Aging in collaboration with other agencies of the United States Public Health Service \citep{h2020}. The study aimed to explore the effect of various clinical, nutritional, \tcr{demographic} and behavioral factors on medical outcomes including morbidity and mortality. The data %,
\tcb{\tcr{were} collected through a baseline visit in 1971 and a follow-up visit in 1982}, and \tcr{the subset we focused on contained data on a cohort of 1425 individuals. A} detailed description of the study data is available at \texttt{https://hsph.harvard.edu/miguel-hernan/causal-inference-book}.

\tcb{Among the various \tcr{bio}medical outcomes recorded in the data, we are interested in the cohort's body weight \tcr{at follow-up}. Based on similar \tcr{data},  %data set
	\citet{ertefaie2020nonparametric} considered the relationship between body weight and smoking from a causal perspective, estimating the average treatment effect of smoking cessation on weight gain \tcr{during the follow up period}. Unlike their study, which divided the observations into two groups according to \tcr{smoking status (quit or not) at follow-up,}
	we analyze all the 1,425 individuals together and our goal is to estimate the median weight $\vt$ of the whole cohort in 1982. Further, we want to explore if there is \tcr{any} significant weight change among the cohort between 1971 and 1982, via comparing the analysis results to the baseline measure, i.e., the median weight 69.40, with a 95\% confidence interval $(68.32,70.48)$, of the\tcr{se} 1,425 individuals in 1971.
	Apart from body weight as the response, we also take \tcr{into} account %of
	$p=20$ important covariates\tcr{, to be included} in our \tcr{imputation} model\tcr{s}, whose names and descriptions are listed in Table \ref{sm_table_data_analysis} of Appendix \ref{sm_data_analysis}. To illustrate our approach in a semi-supervised setup, we randomly select $n=100$ or $200$ out of the 1,425 observations as the labeled data $\cl$ and regard the rest as the unlabeled data $\cu$. Then we implement the supervised and semi-supervised strategies of estimation and inference \tcr{as} described in Section \ref{secs}. Here\tcr{,} regularization is applied to \tcr{all} the regression procedures involved \tcr{in our imputations}. This process is replicated 500 times. \tcr{We take t}he %THE
	median weight $\hvtgs=72.12$, with \tcr{an} estimated standard error \tcr{of} 0.54, of all the 1,425 individuals in 1982 %serves
	as a \tcr{\emph{gold standard}} estimator. %Here the
	\tcm{The subscript ``GS'' in $\hvtgs$ stands for ``gold standard''.} Summarized from the $500$ replications, Table \ref{table_data_analysis} reports the averages of the point estimates and 95\%}

\begin{table}[H]
	\def~{\hphantom{0}}
	\caption{\tcb{Data analysis results of Section \ref{secda}: Estimation and inference of the cohort's median weight in 1982 based on various methods, and efficiencies of the semi-supervised estimators relative to the supervised estimators, i.e., $\E\{(\hvts-\hvtgs)^2\}/\E\{(\hvtss-\hvtgs)^2\}$. The \textbf{boldface} in each case represents the best efficiency and the shortest confidence interval\tcr{.}}
		%{\color{magenta}\bf defined as ????}\tcr{**See comment on the last page -- AC.}
	}
	\vskip 2mm
	
	{
		\resizebox{\textwidth}{!}{
			\begin{tabular}{l|cccccc||cccccc}
				\hline
				\hline
				& \multicolumn{6}{c||}{$n=100$}                                            & \multicolumn{6}{c}{$n=200$}                                            \\
				& Est   & 95\% CI                  & RE            & ESE  & Bias  & CR   & Est   & 95\% CI                  & RE            & ESE  & Bias  & CR   \\
				\hline
				Sup    & 71.85 & (69.77,\,73.93)          & 1.00          & 1.93 & -0.27 & 0.96 & 71.78 & (70.34,\,73.22)          & 1.00          & 1.41 & -0.34 & 0.95 \\
				SS-RF     & 72.14 & (70.61,\,73.68)          & 1.88          & 1.42 & 0.02  & 0.97 & 72.07 & (71.02,\,73.13)          & 2.06          & 1.01 & -0.05 & 0.96 \\
				SS-KS$_1$ & 72.15 & \textbf{(70.74,\,73.56)} & \textbf{2.16} & 1.33 & 0.03  & 0.97 & 72.11 & \textbf{(71.12,\,73.11)} & \textbf{2.57} & 0.90 & -0.01 & 0.96 \\
				SS-KS$_2$ & 72.17 & (70.61,\,73.73)          & 1.71          & 1.49 & 0.05  & 0.96 & 72.10 & (71.07,\,73.13)          & 2.33          & 0.95 & -0.02 & 0.96 \\
				SS-PR     & 72.14 & (70.69,\,73.59)          & 2.01          & 1.37 & 0.02  & 0.97 & 72.11 & (71.10,\,73.12)          & 2.39          & 0.94 & -0.01 & 0.96 \\
				\hline
	\end{tabular}}}
	\label{table_data_analysis}
	
	\begin{footnotesize}
		\tcr{\underline{Glossary of notation}:} $n$, the labeled data size; Est, point estimate; CI, confidence interval; RE, relative efficiency; ESE, empirical standard error; CR, coverage rate of the 95\% confidence intervals; Sup, supervised estimator; SS, semi-supervised estimator; RF, random forest; KS$_1$/KS$_2$, kernel smoothing on the one/two direction(s) selected by linear regression/sliced inverse regression; PR, parametric regression.
	\end{footnotesize}
\end{table}
	\vspace{-0.1in}
	
\noindent \tcb{confidence intervals, and the relative efficiencies $\E\{(\hvts-\hvtgs)^2\}/\E\{(\hvtss-\hvtgs)^2\}$ of the semi-supervised estimators \tcr{versus} % to
	the supervised estimators. We also present the inference results on $\vt$, where the biases and coverage rates are calculated \tcr{relative to} %with
	the gold standard\tcr{,} %estimator
	$\hvtgs$.}

\tcb{In Table \ref{table_data_analysis}, which displays the analysis results on the cohort's median weight in 1982, we notice \tcr{that} the lower bounds of all the semi-supervised confidence intervals are clearly above the upper bound of the 95\% confidence interval $(68.32,70.48)$ of the median weight in 1971, indicating significant weight gain among the cohort between 1971 and 1982. However, this finding is %very
	\tcr{likely} to be ignored by the supervised method \tcr{be}cause the two supervised confidence intervals in Table \ref{table_data_analysis} both overlap \tcr{with} $(68.32,70.48)$. This contrast demonstrates the \tcr{considerable} %remarkable
	advantage of our semi-supervised inference procedures in terms of \tcr{being more powerful in} detecting significance. \tcr{This is also reflected in their efficiencies,} %which are considerably better than the are significantly higher than 1, with best efficiencies of Moreover, we see
	with the various semi-supervised estimators all yield\tcr{ing} \tcr{substantially} better efficiencies \tcr{(with relative efficiencies as high as 2.16 and 2.57, for the two choices of $n$) outperforming} %than %does
	\tcr{the} supervised method. %and
	\tcr{Moreover, they all have negligible biases (much lower than the supervised method) and also} generate satisfactory coverage rates around the nominal level 0.95. These results confirm again the superiority of our semi-supervised method.}

\vspace{-0.01in}
\section{Discussion}\label{sec_discussion}
%For quantile estimation in semi-supervised and high dimensional settings, we have developed a robust and efficient strategy.
\vspace{-0.01in}
\tcr{We considered semi-supervised inference for quantile estimation in high dimensional settings, a problem relatively unaddressed in the existing literature, and developed a robust and efficient strategy %high dimensional settings and also
	based on imputation, allowing for flexible choices of the (nuisance) imputation model. We provided a complete characterization of the achievable estimators, and their robustness and efficiency properties.} Moreover, we %have
\tcmAC{considered} kernel smoothing estimators, \tcr{with possible use of dimension reduction,} as an \tcr{illustration} %instance
of the nuisance estimators involved in our method, establishing novel %{\color{magenta}\bf really? Such results are very tricky, hard to understand, and require stringent conditions.}\tcb{**See comment on next page**}
results \tcr{on} %of
their uniform convergence rates \tcr{in high dimensions}. In a recent parallel work \citep{chakrabortty2022general}, we have also extended the theory and methodology developed in this article to the case of causal parameters. However, the problem of marginal quantiles itself without bringing in the causal framework is of general interest to the broader statistical community, so we focus on it in the current paper.

A natural extension
%{\color{magenta}\bf make sure you think about what you write next. It looks fishy to me.}\tcb{**See comment on next page**}
of \tcr{the} marginal quantile estimation \tcr{problem} is quantile regression \citep{k2005} targeting the $(p+1)$-dimensional parameter $\bbeta_0$ in a possibly misspecified {\it working} model that assumes the $\tau$\tcr{-}conditional quantile of $Y$ given $\X$ equals $(1,\X\trans)\bbeta_0$. One may expect \tcr{that} the one-step update strategy in Section \ref{sec_one_step} with suitable modifications can provide a family of semi-supervised estimators for $\bbeta_0$ that outperform the supervised counterpart. Such a \tcr{procedure,} %process,
however, \tcr{also} involves technical difficulties such as estimating the conditional density of $Y-(1,\X\trans)\bbeta_0$ given $\X$, \tcr{which can be} a \tcr{challenging}
%daunting
task when $p$ is moderate or large, \tcr{and therefore requires developing more careful and sophisticated methodology}. We thus leave this topic for future study.

\phantomsection
\addcontentsline{toc}{section}{Acknowledgements}

\section*{Acknowledgment}
\tcr{AC's research was supported in part by the National Science Foundation grant NSF DMS-2113768.} \tcr{RJC's research was} supported in part by a National Cancer Institute grant and by a Tripods grant from the National Science Foundation.  \tcr{RJC is also affiliated with the School of Mathematical and Physical Sciences, University of Technology Sydney, Broadway NSW 2007, Australia.}

%\abhishek{Ray, like we discussed today, I have added your second affiliation here in the Acknowledgement section and removed it from the front page to save some space there. And I have also added my NSF grant support here.}

\phantomsection
\addcontentsline{toc}{section}{Supplementary Material}

\section*{Supplementary material}
\label{SM}
\begin{itemize}
	\item \tcm{Appendices \ref{sm_technical}--\ref{sm_data_analysis} include
		%an alternative interpretation for the construction of our estimator (Section \ref{sm_alternative_interpretation}),
		technical assumptions required by Theorem \ref{thhd} (Appendix \ref{sm_assumptions}), auxiliary lemmas that would be used for proving the main theorems (Appendix \ref{sm_lemmas}), proofs of all the theoretical results (Appendices \ref{proof_l0}--\ref{proof_thbn}), additional simulation \tcr{results} %outcomes
		(Appendix \ref{sm_additional_simulations}) and \tcr{further information regarding} %descriptions of
		the data analysis in Section \ref{secda} (Appendix \ref{sm_data_analysis}).}
	
	\item  \tcb{All the computer programs \tcr{used for obtaining the results} in Sections \ref{secs}--\ref{secda} are available at: \texttt{https://github.com/guorongdai/semi-supervised\_quantile\_estimation}. }
\end{itemize}

\begin{appendix}
	
	\section{Technical details}\label{sm_technical}
	
	\subsection{Smoothness conditions for kernel smoothing}\label{sm_assumptions}
	The following Assumption  \ref{akernel} contains the %\tcr{basic}
	smoothness conditions required by Theorem \ref{thhd}. These conditions are fairly standard for kernel-based approaches\tcr{, and their analogous versions} %while their analogues
	can \tcr{also} be found in \tcr{various existing works in} the literature, such as \citet{newey1994large}, \citet{andrews1995nonparametric}, \citet{masry1996multivariate} and  \citet{hansen2008uniform}\tcr{, among others}.
	\begin{assumption}\label{akernel}
		(i) The function $K(\cdot)$ is a symmetric kernel of order $d\geq 2$ with a finite $d$th moment. Moreover, it  is bounded, integrable and continuously differentiable. In addition, there exists some constant $v> 1$ such that $\|\nabla K(\s)\|\leq c_1\,\|\s\|^{-v}$ for any $\|\s\|>c_2$. (ii) The support $\ms$ of $\S\equiv\mbP_0\trans\X$ is compact. The density function $f_{\S}(\cdot)$ of $\S$ is bounded and bounded away from zero on $\ms$. Further, it is $d$ times continuously differentiable with a bounded $d$th derivative on $\ms_0$. (iii) With respect to $\s$, the conditional distribution function $F(\theta\mid\S=\s)$ of $Y$ given $\S=\s$ is $d$ times continuously differentiable and has a bounded $d$th derivatives on $\ms_0\times\mbtv$.
	\end{assumption}
	
	%\abhishek{Need some commenting and justification, along with citations, for this assumption.}
	
	\subsection{Preliminary lemmas}\label{sm_lemmas}
	The following Lemmas \ref{l0}--\ref{1v2} would be useful in the proofs of the main theorems and propositions. The proofs of these lemmas, as well as Theorems \ref{thos}--\ref{thhd} and Propositions \ref{thphi}--\ref{thbn}, can be found in Sections \ref{proof_l0}--\ref{proof_thbn}.
	\begin{lemma}\label{l0}
		For some fixed integer $M$, suppose $W_{n,1},\ldots,W_{n,M}\in\rR$ are mutually independent sequences of random variables satisfying that, for some constants $\mu_m$ and $\sigma_m>0$,
		\be
		W_{n,m}\xrightarrow{d}\mn(\mu_m,\sigma^2_m)\quad (m=1,\ldots,M;n\to\infty).
		\label{cid}
		\ee
		Then\tcmAC{,} $\sm W_{n,m}\xrightarrow{d} \mn(\sm\mu_m,\sm\sigma^2_m)$ $(n\to\infty)$.
	\end{lemma}
	
	\begin{lemma}\label{1v2}
	Suppose there are two independent samples, $\ms_1$ and $\ms_2$, consisting of $n$ and $m$ independent copies of $(\X\trans,Y)\trans$, respectively. For $\bgamma\in\rR^d$ with some fixed $d$, let $\hg_{n}(\x,\bgamma)$ be an estimator of a measurable function $g(\x,\bgamma)\in\rR$ based on $\ms_1$ and
	\bse
	\mbG_{m}\{\hg_{n}(\X,\bgamma)\}:= m^{1/2}[m^{-1}\hbox{$\sum_{(\X_i\trans,Y_i)\trans\in\ms_2}$}\hg_{n}(\X_i,\bgamma)-\E_\X\{\hg_{n}(\X,\bgamma)\}].
	\ese
	For some set $\ct\subset\rR^d$, denote
	\bse
	\Delta(\ms_1):=(\sg\E_\X[\{\hg_n(\X,\bgamma)\}^2])^{1/2}\hbox{ \tcm{and} }   M(\ms_1):=\sgx|\hg_n(\x,\bgamma)|.
	\ese
	For any $\eta\in(0,\Delta(\ms_1)+c\,]$, suppose $\G_{n}:=\{\hg_{n}(\X,\bgamma):\bgamma\in\ct\}$ satisfies that
	\be
	N_{[\,]}\{\eta,\G_{n}\mid\ms_1,L_2(\P_\X)\}\leq H(\ms_1)\eta^{-c} \tcmAC{,}
	\label{bracket2}
	\ee
	with some function $H(\ms_1)>0$. Here $\G_n$ is indexed by $\bgamma$ only and treats $\hg_n(\cdot,\bgamma)$ as a nonrandom function. Assume $H(\ms_1)=O_p(a_n)$, $\Delta(\ms_1)=O_p(d_{n,2})$ and $M(\ms_1)=O_p(d_{n,\infty})$  with some positive sequences $a_n$, $d_{n,2}$ and $d_{n,\infty}$ allowed to diverge\tcmAC{. Then,} we have\tcmAC{:}
	\bse
	\sg|\mbG_m\{\hg_n(\X,\bgamma)\}|=O_p(r_{n,m}),
	\ese
	where $r_{n,m}=d_{n,2}\{\log\,a_n+\log\,(d_{n,2}^{-1})\}+m^{-1/2}d_{n,\infty}\{(\log\,a_n)^2+(\log\,d_{n,2})^2\}$.
\end{lemma}

\subsection{Proof of Lemma \ref{l0}}\label{proof_l0}
We have that, for any $t\in\rR$,
\be
\E\{\exp(it\,\sm W_{n,m})\}&=&\hbox{$\prod_{m=1}^M$}\E\{\exp(it\, W_{n,m})\} \nonumber\\
&\to&\hbox{$\prod_{m=1}^M$}\exp(i\mu_m t-\sigma^2_m t^2/2)  \nonumber\\
&=&\exp\{i(\sm\mu_m )t-(\sm\sigma^2_m) t^2/2\} \label{cf}
\ee	
where $i$ is the imaginary unit. In the above, the first step uses the mutual independence of $W_{n,1},\ldots,W_{n,M}$, and the second step is due to (\ref{cid}) and Levy's continuity theorem. The fact that (\ref{cf}) is the characteristic function of  $\mn(\sm\mu_m,\sm\sigma^2_m)$ implies the conclusion.

\hfill$\square$

\subsection{Proof of Lemma \ref{1v2}}\label{proof_lv2}
For any $\delta\in(0,\Delta(\ms_1)+c\,]$, we have that the bracketing integral\tcmAC{:}
\bse
J_{[\,]}\{\delta,\G_n\mid\ms_1,L_2(\P_\X)\}~ &\equiv&~\hbox{$\int_0^\delta$}[1+\log\,N_{[\,]}\{\eta,\G_n\mid\ms_1,L_2(\P_\X)\}]^{1/2}d\eta \\
&\leq&~\hbox{$\int_0^\delta$}1+\log \,N_{[\,]}\{\eta,\G_n\mid\ms_1,L_2(\P_\X)\}d\eta  \\
&\leq&~\hbox{$\int_0^\delta$}1+\log\,H(\ms_1)-c\,\log\,\eta\, d\eta \\
&=&~\delta\{1+\log\,H(\ms_1)\}+c\,(\delta-\delta\,\log\,\delta),
\ese
where the third step is due to (\ref{bracket2}). This, combined with Lemma 19.36 of \citet{van2000asymptotic}, implies that
\bse
&&\phantom{=}\E_\X[\sg|\mbG_m\{\hg_n(\X,\bgamma)\}|] \\
&&\leq J_{[\,]}\{\delta,\G_n\mid\ms_1,L_2(\P_\X)\}+[J_{[\,]}\{\delta,\G_n\mid\ms_1,L_2(\P_\X)\}]^2M(\ms_1)\delta^{-2}m^{-1/2} \\
&&\leq \delta\{1+\log\,H(\ms_1)\}+c\,(\delta-\delta\,\log\,\delta)+\{1+\log\,H(\ms_1)+c\,(1-\log\,\delta)\}^2M(\ms_1)m^{-1/2}
\ese
for any $\delta\in(\Delta(\ms_1),\Delta(\ms_1)+c\,]$. Therefore\tcmAC{,}
\bse
\E_\X[\sg|\mbG_m\{\hg_n(\X,\bgamma)\}|] &\leq& \Delta(\ms_1)\{1+\log\,H(\ms_1)\}+c\,\{\Delta(\ms_1)-\Delta(\ms_1)\,\log\,\Delta(\ms_1)\}+\\
&&[1+\log\,H(\ms_1)+c\,\{1-\log\,\Delta(\ms_1)\}]^2M(\ms_1)m^{-1/2}.
\ese
Since the right hand side in the above is $O_p(r_{n,m})$, it gives that
\be
\E_\X[\sg|\mbG_m\{\hg_n(\X,\bgamma)\}|] =O_p(r_{n,m}).
\label{ex}
\ee
Then, for any positive sequence $t_n\to\infty$, we have\tcmAC{:}
\bse
&&\phantom{=}\pr[\sg|\mbG_m\{\hg_n(\X,\bgamma)\}|>t_n r_{n,m}\mid\ms_1] \\
&&\leq (t_n r_{n,m})^{-1}\E_\X[\sg|\mbG_m\{\hg_n(\X,\bgamma)\}|] =o_p(1),
\ese
where the first step holds by Markov's inequality and the last step is due to (\ref{ex}). This, combined with Lemma 6.1 of \citet{chernozhukov2018double}, gives that
\bse
\pr[\sg|\mbG_m\{\hg_n(\X,\bgamma)\}|>t_n r_{n,m}]\to 0,
\ese
which completes the proof. \hfill{$\square$}

\subsection{Proof of Theorem \ref{thos}}\label{proof_thos}
Write
\be
\hvtss-\vt=\{S_1(\hvti)-\vt\}+\{\hf(\hvti)\}^{-1}\{S_2(\hvti)+S_3(\hvti)\},
\label{formulate}
\ee
where
\bse
&&S_1(\theta):=\theta-\{\hf(\theta)\}^{-1}\E_n\{\psi(Y,\theta)\},\\
&&S_2(\theta):=(1-\nu_{n,N})[\E_n\{\hphi(\X,\theta)-\phi(\X,\theta)\}-\E_N\{\hphi(\X,\theta)-\phi(\X,\theta)\}],\\
&&S_3(\theta):=\E_n\{\phi(\X,\theta)\}-\E_{n+N}\{\phi(\X,\theta)\}.
\ese

Assumption \ref{ainit} gives\tcmAC{:}
\be
&&\pr\{\hvti\in\mbtv\}\to 1, \label{belong}\\
&&\hL:=\{\hf(\hvti)\}^{-1}-\{f(\vt)\}^{-1}=O_p(v_n)=o_p(1).
\label{hl}
\ee

By Theorem 19.3 of \citet{van2000asymptotic}, we know that $\{I(Y<\theta):\theta\in\mbtv\}$ forms a $\P$-Donsker class, so the permanence properties of $\P$-Donsker classes \citep{van1996weak} guarantee that
\be
\md:=\{\psi(Y,\theta):\theta\in\mbtv\}=\{I(Y<\theta)-\tau:\theta\in\mbtv\}  \label{donsker}
\ee
is also a $\P$-Donsker class. Moreover, the convergence (\ref{belong}) implies that $\psi(Y,\hvti)$ is in $\md$ with probability tending to one. In addition, we have\tcmAC{:}
\bse
\E[\{\psi(Y,\hvti)-\psi(Y,\vt)\}^2]&=& \E[\{I(Y<\hvti)-I(Y<\vt)\}^2] \\
&=&F(\hvti)+F(\vt)-2F\{\min(\hvti,\vt)\}\to 0.
\ese
in probability because of the continuity of $F(\cdot)$ from Assumption \ref{adensity} and the consistency of $\hvti$ from Assumption \ref{ainit}. Hence Lemma 19.24 of \citet{van2000asymptotic} gives that
\be
\mbG_n\{\psi(Y,\hvti)-\psi(Y,\vt)\}=o_p(1),
\label{mbg}
\ee
which implies that
\be
\E_n\{\psi(Y,\hvti)\}-\E\{\psi(Y,\hvti)\} &=&n^{-1/2}[\mbG_n\{\psi(Y,\vt)\} +\mbG_n\{\psi(Y,\hvti)-\psi(Y,\vt)\}] \nonumber\\
&=&\E_n\{\psi(Y,\vt)\}+o_p(n^{-1/2}).
\label{df11}
\ee
Taylor's expansion gives that
\be
\E\{\psi(Y,\hvti)\}&=&f(\vt)(\hvti-\vt)+O_p(|\hvti-\vt|^2)\nonumber\\
&=&f(\vt)(\hvti-\vt)+O_p(u_n^2) \label{df12} \\
&=&O_p(u_n),\label{df122}
\ee
where the residual term in the first step is due to (\ref{belong}) as well as the fact that $f(\cdot)$ has a bounded derivative in $\mbtv$ from Assumption \ref{adensity}, the second step uses Assumption \ref{ainit}, and the last step holds by the fact that $u_n=o(1)$ from Assumption \ref{ainit}. Then\tcmAC{,} we have\tcmAC{:}
\be
\phantom{=}\hL\E_n\{\psi(Y,\hvti)\}&=&\hL[\E_n\{\psi(Y,\vt)\}+O_p(u_n)+o_p(n^{-1/2})] \nonumber\\
&=&\hL\{O_p(n^{-1/2})+O_p(u_n)+o_p(n^{-1/2})\} \nonumber\\
&=&O_p(u_nv_n)+o_p(n^{-1/2}),
\label{df13}
\ee
where the first step holds by (\ref{df11}) and (\ref{df122}), the second step uses the central limit theorem and the last step is due to (\ref{hl}). Thus\tcmAC{,} we have\tcmAC{:}
\be
&&\phantom{=}S_1(\hvti)-\vt \nonumber\\
&&=\hvti-\vt-\{\hf(\hvti)\}^{-1}\E_n\{\psi(Y,\hvti)\} \nonumber\\
&&=\hvti-\vt-\{f(\vt)\}^{-1}\E_n\{\psi(Y,\hvti)\}+O_p(u_nv_n)+o_p(n^{-1/2}) \nonumber\\
&&=\hvti-\vt-\{f(\vt)\}^{-1}[\E\{\psi(Y,\hvti)\}+\E_n\{\psi(Y,\vt)\}]+O_p(u_nv_n)+o_p(n^{-1/2}) \nonumber\\
&&=-\{f(\vt)\}^{-1}\E_n\{\psi(Y,\vt)\}+O_p(u_n^2+u_nv_n)+o_p(n^{-1/2}),
\label{ex1}
\ee
where the second step uses (\ref{df13}), the third step holds by (\ref{df11}) and the last step is due to (\ref{df12}).

Moreover, denote
\bse
\mbG_{n_\kK,k}\{\hpsi(\X,\theta)\}=n_\kK^{1/2}[n_\kK^{-1}\hbox{$\sum_{i\in\I_k}$}\hpsi(\X_i,\theta)- \E_\X\{\hpsi(\X,\theta)\}]\quad (k=1,\ldots,\kK).
\ese
Considering Assumption \ref{aest}, Lemma \ref{1v2} gives that
\be
&&\phantom{=}\sb|\mbG_{n_\kK,k}\{\hpsi(\X,\theta)\}|=O_p(r_n), \nonumber\\
&&\phantom{=}\sb|\mbG_N\{\hpsi(\X,\theta)\}| \nonumber\\
&&=O_p[d_{n,2}\{\log\,a_n+\log\,(d_{n,2}^{-1})\}+N^{-1/2}d_{n,\infty}\{(\log\,a_n)^2+(\log\,d_{n,2})^2\}] \nonumber\\
&&=O_p(r_n)\quad (k=1,\ldots,\kK). \label{rr2}
\ee
Hence, using (\ref{belong}), we have that, with probability tending to one,
\be
|S_2(\hvti)|&\leq&\sb|\E_n\{\hphi(\X,\theta)-\phi(\X,\theta)\}-\E_N\{\hphi(\X,\theta)-\phi(\X,\theta)\}| \nonumber\\
&=&\sb|\kK^{-1}\sk[n_\kK^{-1/2}\mbG_{n_\kK,k}\{\hpsi(\X,\theta)\}- \nonumber\\
&&\phantom{\sb|\kK^{-1}\sk[}N^{-1/2}\mbG_N\{\hpsi(\X,\theta)\}]|\nonumber \\
&\leq& \kK^{-1}\sk [n_\kK^{-1/2}\sb|\mbG_{n_\kK,k}\{\hpsi(\X,\theta)\}|+\nonumber\\
&&\phantom{\kK^{-1}\sk [}N^{-1/2}\sb|\mbG_N\{\hpsi(\X,\theta)\}|]=O_p(n^{-1/2}r_n).
\label{s2int}
\ee
In addition, we know
\be
\hf(\hvti)=O_p(1)
\label{hfo}
\ee
due to the facts that $\hf(\hvti)-f(\vt)=o_p(1)$ from Assumption \ref{ainit}, and that $f(\vt)>0$ from Assumption \ref{adensity}. Combining (\ref{s2int}) and (\ref{hfo}) yields\tcmAC{:}
\be
\{\hf(\hvti)\}^{-1}S_2(\hvti)=O_p(n^{-1/2}r_n).
\label{ex2}
\ee

Next, we have that
\be
S_3(\hvti)&=&(\E_n-\E_{n+N})\{\phi(\X,\hvti)\} \nonumber\\
&=&(\E_n-\E_{n+N})\{\phi(\X,\vt)\}+n^{-1/2}\mbG_n\{\phi(\X,\hvti)-\phi(\X,\vt)\}-\nonumber \\
&&(n+N)^{-1/2}\mbG_{n+N}\{\phi(\X,\hvti)-\phi(\X,\vt)\}\nonumber \\
&=&(\E_n-\E_{n+N})\{\phi(\X,\vt)\}+o_p(n^{-1/2})+o_p\{(n+N)^{-1/2}\} \nonumber \\
&=&(\E_n-\E_{n+N})\{\phi(\X,\vt)\}+o_p(n^{-1/2})
\label{s1}
\ee
where the third step uses (\ref{belong}) and (\ref{uni1}) in Assumption \ref{aimp}. Therefore\tcmAC{,} it follows that
\be
\hL S_3(\hvti)&=&\hL[n^{-1/2}\mbG_n\{\phi(\X,\vt)\}-(n+N)^{-1/2}\mbG_{n+N}\{\phi(\X,\vt)\}+o_p(n^{-1/2})] \nonumber\\
&=&o_p(n^{-1/2}),
\label{s2}
\ee
where the last step holds by (\ref{hl}) as well as the fact that $\mbG_n\{\phi(\X,\vt)\}=O_p(1)$ and $\mbG_{n+N}\{\phi(\X,\vt)\}=O_p(1)$ ensured by the central limit theorem and the square integrability of $\phi(\X,\vt)$ from Assumption \ref{aimp}. Combining (\ref{s1}) and (\ref{s2}) yields\tcmAC{:}
\be
\{\hf(\hvti)\}^{-1}S_3(\hvti)&=&\{f(\vt)\}^{-1}S_3(\hvti)+o_p(n^{-1/2}) \nonumber\\
&=&\{f(\vt)\}^{-1}(\E_n-\E_{n+N})\{\phi(\X,\vt)\}+o_p(n^{-1/2})
\label{ex3}.
\ee

Summing up, the equations (\ref{formulate}), (\ref{ex1}), (\ref{ex2}) and (\ref{ex3}) imply that
\be
\quad\hvtss-\vt=\{f(\vt)\}^{-1}\E_n\{\omega_{n,N}(\Z,\vt)\}+O_p(u_n^2+u_nv_n+n^{-1/2}r_n)+o_p(n^{-1/2}).
\qquad \label{expansion}
\ee

Further, we know that
\be
n^{1/2}\E_n\{\omega_{n,N}(\Z,\vt)\}&=&\mbG_n\{(1-\nu_{n,N})\phi(\X,\vt)-\psi(Y,\vt)\}- \nonumber\\
&&(n/N)^{1/2}\mbG_N\{(1-\nu_{n,N})\phi(\X,\vt)\}.
\label{sum}
\ee
The central limit theorem and Slutsky's theorem give that, as $n,N\to\infty$,
\be
&&\sigma_1^{-1}\mbG_n\{(1-\nu_{n,N})\phi(\X,\vt)-\psi(Y,\vt)\}\to\mn(0,1), \label{clt1}\\
&&\sigma_2^{-1}(n/N)^{1/2}\mbG_N\{(1-\nu_{n,N})\phi(\X,\vt)\}\to\mn(0,1), \label{clt2}
\ee
where
\bse
&&\sigma_1^2:=\E[\{\psi(Y,\vt)\}^2]+(1-\nu_{n,N})^2\var\{\phi(\X,\vt)\}-2(1-\nu_{n,N})\E\{\psi(Y,\theta)\phi(\X,\theta)\}, \\
&&\sigma_2^2:=(n/N)(1-\nu_{n,N})^2\var\{\phi(\X,\vt)\}.
\ese
Thus\tcmAC{,} we have\tcmAC{:}
\be
\sigma_1^2+\sigma_2^2&=&\E[\{\psi(Y,\vt)\}^2]+(1-\nu_{n,N})\var\{\phi(\X,\vt)\}-2(1-\nu_{n,N})\E\{\phi(\X,\vt)\psi(Y,\vt)\} \nonumber\\
&=&(1-\nu_{n,N})\var\{\psi(Y,\vt)-\phi(\X,\vt)\}+\nu_{n,N}\var\{\psi(Y,\vt)\}=\sigss^2.
\label{sigma}
\ee
Finally, applying Lemma \ref{l0} and Slutsky's theorem, the equations (\ref{expansion})--(\ref{sigma}) conclude the asymptotic normality under the assumption (\ref{negligible}) and the independence of the empirical processes in (\ref{clt1}) and (\ref{clt2}). \hfill{$\square$}

\subsection{Proof of Corollary \ref{cor1}}\label{proof_cor1}
Since $\nu=0$, the central limit theorem gives that
\bse
\E_{n+N}\{\phi(\X,\vt)\}=\E\{\phi(\X,\vt)\}+O_p\{(n+N)^{-1/2}\}=\E\{\phi(\X,\vt)\}+o_p(n^{-1/2}).
\ese
This, combined with (\ref{expansion}), implies the stochastic expansion, followed by the asymptotic normality. Further, it is clear that
$\sigss\to\tsigss$ as $n\to\infty$ in that $\lim_{n\to\infty}\nu_{n,N}=0$. \hfill{$\square$}

\subsection{Proof of Proposition \ref{thphi}}\label{proof_thphi}
The second moment of $\phi(\X,\vt)$ is obviously finite because the function $F(\cdot\mid\S)$ is bounded. For any $\theta_1,\theta_2\in\mbtv$, Taylor's expansion gives\tcmAC{:}
\bse
|\phi(\X,\theta_1)-\phi(\X,\theta_2)|&=&|F(\theta_1\mid\S)-F(\theta_2\mid\S)| \\
&\leq&\sb f(\theta\mid\S)|\theta_1-\theta_2|,
\ese
Therefore, under the assumption that
\be
\E[\{\sb f(\theta\mid\S)\}^2]<\infty,
\label{aphi}
\ee
Example 19.7 of \citet{van2000asymptotic} implies\tcmAC{:}
\be
N_{[\,]}\{\eta,\mathcal{F},L_2(\P_\X)\}\leq c\,\eta^{-1}
\label{fstar}
\ee
with $\mathcal{F}:=\{\phi(\X,\theta):\theta\in\mbtv\}$. Then, by Theorem 19.5 of \citet{van2000asymptotic}, we know that $\mathcal{F}$ is $\P$-Donsker. Further, we have that, for any sequence $\tvt\to\vt$ in probability,
\bse
\E_\X[\{\phi(\X,\tvt)-\phi(\X,\vt)\}^2]&=&\E_\S[\{F(\tvt\mid\S)-F(\vt\mid\S)\}^2] \\
&\leq&(\tvt-\vt)^2\E[\{\sb f(\theta\mid\S)\}^2]\to 0
\ese
in probability, where the second step uses Taylor's expansion and the fact that $\tvt\in\mbtv$ with probability approaching one, and the last step holds by (\ref{aphi}). Lastly, applying Lemma 19.24 of \citet{van2000asymptotic} concludes (\ref{uni1}).  \hfill{$\square$}

\subsection{Proof of Theorem \ref{thhd}}\label{proof_thhd}
Set $m(\x,\theta,\mbP):=\phi(\x,\theta,\mbP)f_\S(\mbP\trans\x)$. We now derive the convergence rate of $\hm(\x,\theta,\hmbP)-m(\x,\theta,\mbP_0)$.

We first handle the error from estimating $\mbP_0$ by $\hmbP$, i.e., $\hm(\x,\theta,\hmbP)-\hm(\x,\theta,\mbP_0) $. Taylor's expansion gives that, for
\be
\bar{\s}_n:=h_n^{-1}\{\mbP_0\trans+\bmu(\hmbP-\mbP_0)\trans\}(\x-\X)
\label{sbar}
\ee
with some $\bmu:=\diag(\mu_1,\ldots,\mu_r)$ and $\mu_j\in(0,1)$ $(j=1,\ldots,r)$,
\be
&&\phantom{=}\hm(\x,\theta,\hmbP)-\hm(\x,\theta,\mbP_0) \nonumber \\
&&=h_n^{-(r+1)}\Enk[\{\nabla K(\bar{\s})\}\trans(\hmbP-\mbP_0)\trans(\x-\X)\psi(Y,\theta)] \nonumber\\
&&= h_n^{-(r+1)}\trace ((\hmbP-\mbP_0)\trans \Enk[(\x-\X)\{\nabla K(\bar{\s})\}\trans\psi(Y,\theta)]) \nonumber\\
&&=h_n^{-(r+1)}\trace[(\hmbP-\mbP_0)\trans\{\bfU_{n,1}(\x,\theta)+\bfU_{n,2}(\x,\theta)-\bfU_{n,3}(\x,\theta)\}],
\label{un}
\ee
where
\bse
&&\bfU_{n,1}(\x,\theta):=\Enk((\x-\X)[\nabla K(\bar{\s}_n)-\nabla K\{h_n^{-1}\mbP_0\trans(\x-\X)\}]\trans\psi(Y,\theta)),  \\
&&\bfU_{n,2}(\x,\theta):=\Enk(\x [\nabla K\{h_n^{-1}\mbP_0\trans(\x-\X)\}]\trans\psi(Y,\theta)),  \\
&&\bfU_{n,3}(\x,\theta):=\Enk(\X [\nabla K\{h_n^{-1}\mbP_0\trans(\x-\X)\}]\trans\psi(Y,\theta)).
\ese
For the function $\rho(\cdot)$ in Assumption \ref{ahbe} (ii), denote $\mathcal{J}_n:=\{h^{-r}_n\rho\{h_n^{-1}(\s-\S)\}:\s\in\ms\}$. Taylor's expansion gives that, for  any $\s_1,\s_2\in\ms$  and some $\bar{\s}:=\s_1+\bmu(\s_2-\s_1)$ with $\bmu:=\diag(\mu_1,\ldots,\mu_r)$ and $\mu_j\in(0,1)$ $(j=1,\ldots,r)$,
\bse
&&\phantom{=}h^{-r}_n|\rho\{h_n^{-1}(\s_1-\S)\}-\rho\{h_n^{-1}(\s_2-\S)\}| \\
&&= h_n^{-(r+1)}|[\nabla\rho\{h_n^{-1}(\bar{\s}-\S)\}]\trans(\s_1-\s_2)|\leq c\,h^{-(r+1)}_n\|\s_1-\s_2\|,
\ese
where the second step uses the boundedness of $\nabla\rho(\cdot)$ from Assumption \ref{ahbe} (ii). Therefore Example 19.7 of \citet{van2000asymptotic} implies\tcmAC{:}
\be
N_{[\,]}\{\eta ,\mathcal{J}_n,L_2(\P_\X)\}\leq c\,h_n^{-(r+1)}\eta^{-r}.
\label{bracj}
\ee
Moreover, we have that
\be
\sss [h^{-r}_n\rho\{h_n^{-1}(\s-\S)\}]=O(h_n^{-r}).
\label{supj}
\ee
due to the boundedness of $\rho(\cdot)$ from Assumption \ref{ahbe} (ii). In addition, we know that
\be
\sups\E_\S([h_n^{-r}\rho \{h_n^{-1}(\s-\S) \}]^2)&=&h^{-r}\sups\hbox{$\int$}h_n^{-r}[\rho\{h_n^{-1}(\s-\bfv) \}]^2 f_\S(\bfv)d\bfv \nonumber\\
&=&h_n^{-r}\sups\hbox{$\int$}\{\rho(\bft )\}^2 f_\S(\s-h_n\bft)d\bft = O(h_n^{-r}),\quad
\label{varj}
\ee
where the second step uses change of variables and the last step holds by the boundedness of $f_\S(\cdot)$ from Assumption \ref{akernel} (ii) and the square integrability of $\rho(\cdot)$ from Assumption \ref{ahbe} (ii). Based on (\ref{bracj})--(\ref{varj}), applying Lemma \ref{1v2} yields that
\be
&&\phantom{=}\sups|\Enk[h^{-r}_n\rho\{h_n^{-1}(\s-\S)\}]-\E_\X[h^{-r}_n\rho\{h_n^{-1}(\s-\S)\}]| \nonumber\\
&&=O_p\{n_{\kK^-}^{-1/2}h_n^{-r/2}\log(h_n^{-1})+n_{\kK^-}^{-1}h_n^{-r}(\log\,h_n)^2\}=o_p(1),
\label{grho}
\ee
where the second step is because we assume $(nh_n^r)^{-1/2}\log(h_n^{-r})=o(1)$. Then\tcmAC{,} we know
\bse
\sups\E_\S[h_n^{-r}\rho \{h_n^{-1}(\s-\S) \}]&=&\sups\hbox{$\int$}h_n^{-r}\rho\{h_n^{-1}(\s-\bfv) \} f_\S(\bfv)d\bfv  \\
&=&\sups\hbox{$\int$}\rho(\bft ) f_\S(\s-h_n\bft)d\bft = O(1).
\ese
where the second step uses change of variables and the last step holds by the boundedness of $f_\S(\cdot)$ from Assumption \ref{akernel} (ii) and the integrability of $\rho(\cdot)$ from Assumption \ref{ahbe} (ii). This, combined with (\ref{grho}), implies that
\be
\sups\Enk[h_n^{-r}\rho \{h_n^{-1}(\s-\S) \}]=O_p(1).
\label{exrho}
\ee
Next, we have\tcmAC{:}
\be
&&\phantom{=}\sx\Enk [\|\nabla K(\bar{\s}_n)-\nabla K\{h_n^{-1}\mbP_0\trans(\x-\X)\}\|] \nonumber\\
&&\leq\sx\Enk [\|\bar{\s}_n-h_n^{-1}\mbP_0\trans(\x-\X)\|\rho\{h_n^{-1}\mbP_0\trans(\x-\X)\}] \nonumber\\
&&\leq\sx\Enk [\|(\hmbP-\mbP_0)\trans(\x-\X)\|h_n^{-1}\rho\{h_n^{-1}\mbP_0\trans(\x-\X)\}] \nonumber\\
&&\leq c\,\|\hmbP-\mbP_0\|_1\sxx\|\x-\X\|_{\infty}\sups\Enk [h_n^{-1}\rho\{h_n^{-1}(\s-\S)\}] \nonumber \\ &&=O_p(h_n^{r-1}\alpha_n),
\label{alphan}
\ee
where the first step uses the local lipschitz continuity of $\nabla K(\cdot)$ from Assumption \ref{ahbe} (ii), the second step is due to the definition (\ref{sbar}) of $\bar{\s}_n$, the third step holds by H\"older's inequality, and the last step is because of Assumptions \ref{al1}, \ref{ahbe} (i) and the equation (\ref{exrho}). Hence\tcmAC{,}
\bse
&&\phantom{=}\sbx\|\bfU_{n,1}(\x,\theta)\|_{\infty} \\
&&\leq c\,\sx\Enk [\|\x-\X\|_{\infty}\|\nabla K(\bar{\s}_n)-\nabla K\{h_n^{-1}\mbP_0\trans(\x-\X)\}\|] \\
&&\leq c\,\sx\Enk [\|\nabla K(\bar{\s}_n)-\nabla K\{h_n^{-1}\mbP_0\trans(\x-\X)\}\|] =O_p(h_n^{r-1}\alpha_n).
\ese
where the first step holds by the boundedness of $\psi(Y,\theta)$, the second step is due to Assumption \ref{ahbe} (i), and the last step uses (\ref{alphan}). This, combined with Assumption \ref{al1} and H\"older's inequality, implies that
\be
&&\phantom{=}\sbx\|(\hmbP-\mbP_0)\trans \bfU_{n,1}(\x,\theta)\|_\infty \nonumber\\
&&\leq\|\hmbP-\mbP_0\|_1\sbx\|\bfU_{n,1}(\x,\theta)\|_{\infty}=O_p(h_n^{r-1}\alpha_n^2).
\label{bdn1}
\ee
Next, under Assumptions \ref{akernel} (ii) and \ref{ahbe} (ii) as long as the fact that $\{\psi(Y,\theta):\theta\in\mbtv\}$ is a VC class with a bounded envelope function, Lemma B.4 of \citet{escanciano2014uniform} gives that
\be
&&\sbx\|\bfU_{n,2}(\x,\theta)-\E\{\bfU_{n,2}(\x,\theta)\}\|_{\infty}=O_p(h_n^{r}\gamma_n), \label{dn2}\\
&&\sbx\|\bfU_{n,3}(\x,\theta)-\E\{\bfU_{n,3}(\x,\theta)\}\|_{\infty}=O_p(h_n^{r}\gamma_n).
\label{dn3}
\ee
Let $\delta(\s,\theta):=\E\{\psi(Y,\theta)\mid\S=\s\}f_\S(\s)$ and $\nabla\delta(\s,\theta):=\partial \delta(\s,\theta)/\partial \s$. We have\tcmAC{:}
\be
&&\phantom{=}\sbx\|\E\{\bfU_{n,2}(\x,\theta)\}\|_\infty \nonumber\\
&&\leq \sbx\|\x\hbox{$\int$}\delta(\s,\theta)[\nabla K\{h_n^{-1}(\mbP_0\trans\x-s)\}]\trans ds\|_\infty \nonumber\\
&&=h^{r+1}\sbx\|\x\hbox{$\int$}\{\nabla\delta(\mbP_0\trans\x-h_n\bft,\theta)\}\trans K(\bft)d\bft\|_\infty =O(h^{r+1}).
\label{edn2}
\ee
In the above, the second step uses integration by parts and change of variables, while the last step holds by Assumption \ref{ahbe} (i), the boundedness of $\nabla\delta(\s,\theta)$ from Assumptions \ref{akernel} (ii) and (iii), as well as the integrability of $K(\cdot)$ from Assumption \ref{akernel} (i). Set $\bzeta(\s,\theta):=f_\S(\s)\bfeta_1(\s,\theta)$ and $\nabla\bzeta(\s,\theta):=\partial \bzeta(\s,\theta)/\partial \s$. Analogous to (\ref{edn2}), we know
\be
&&\phantom{=}\sbx\|\E\{\bfU_{n,3}(\x,\theta)\}\|_\infty  \nonumber\\
&&\leq \sbx\|\hbox{$\int$}\bzeta(\s,\theta) [\nabla K\{h_n^{-1}(\mbP_0\trans\x-s)\}]\trans ds\|_\infty \nonumber\\
&&=h^{r+1}\sbx\|\hbox{$\int$}\{\nabla\bzeta(\mbP_0\trans\x-h_n\bft,\theta)\}\trans K(\bft)d\bft\|_\infty =O(h^{r+1}),
\label{edn3}
\ee
where the last step holds by the boundedness of $\|\nabla\bzeta(\s,\theta)\|_\infty$ from Assumptions \ref{akernel} (ii) and \ref{ahbe} (iii), and the integrability of $K(\cdot)$ from Assumption \ref{akernel} (i). Combining (\ref{dn2})--(\ref{edn3}) yields\tcmAC{:}
\bse
\sbx\|\bfU_{n,2}(\x,\theta)-\bfU_{n,3}(\x,\theta)\|_\infty=O_p(h_n^{r}\gamma_n+h_n^{r+1}),
\ese
which implies that
\bse
&&\phantom{=}\sbx\|(\mbP_0-\hmbP)\trans\{\bfU_{n,2}(\x,\theta)-\bfU_{n,3}(\x,\theta)\}\|_\infty \\
&&\leq\|\mbP_0-\hmbP\|_1\sbx\|\bfU_{n,2}(\x,\theta)-\bfU_{n,3}(\x,\theta)\|_{\infty} \\
&&=O_p(h_n^{r}\gamma_n\alpha_n+h_n^{r+1}\alpha_n)
\ese
using H\"older's inequality and Assumption \ref{al1}. This, combined with (\ref{un}) and (\ref{bdn1}), gives\tcmAC{:}
\be
\sbx|\hm(\x,\theta,\hmbP)-\hm(\x,\theta,\mbP_0)| =O_p(s_{n,2}).
\label{an2}
\ee

Moreover, we control the error $\hm(\x,\theta,\mbP_0)-m(\x,\theta,\mbP_0)$. Under Assumptions \ref{akernel} (i), (ii) and the fact that $\{\psi(Y,\theta):\theta\in\mbtv\}$ is a VC class with a bounded envelope function, Lemma B.4 of \citet{escanciano2014uniform} gives that
\be
\sbx|\hm(\x,\theta,\mbP_0)-\E\{\hm(\x,\theta,\mbP_0)\}|=O_p(\gamma_n).
\label{pt2}
\ee
Further, under Assumption \ref{akernel}, standard arguments based on $d$th order Taylor's expansion of $m(\x,\theta,\mbP_0)$ yield that
\be
\sbx|\E\{\mnk(\x,\theta,\mbP_0)\}-m(\x,\theta,\mbP_0)|=O(h_n^d).
\label{pt3}
\ee

Combining (\ref{an2})--(\ref{pt3}) yields\tcmAC{:}
\be
\sbx|\hm(\x,\theta,\hmbP)-m(\x,\theta,\mbP_0)|=O_p(s_{n,1}+s_{n,2}).
\label{num}
\ee
Similar arguments give that
\be
\sx|\hl(\x,\hmbP)-f_{\S}(\mbP_0\trans\x)|=O_p(s_{n,1}+s_{n,2}).
\label{deno}
\ee
Thus\tcmAC{,} we have\tcmAC{:}
\bse
&&\phantom{=}\sbx|\hphik(\x,\theta,\hmbP)-\phi(\x,\theta,\mbP_0)| \nonumber\\
&&=\sbx|\{\hl(\x,\hmbP)\}^{-1}\hm(\x,\theta,\hmbP)-\{\ell(\x,\mbP_0)\}^{-1}m(\x,\theta,\mbP_0)| \\
&&\leq\sbx|\{\hl(\x,\mbP_0)\}^{-1}\{\hm(\x,\theta,\hmbP)-m(\x,\theta,\mbP_0)\}|+ \\
&&\phantom{=}\sbx|[\{\hl(\x,\mbP_0)\}^{-1}-\{\ell(\x,\mbP_0)\}^{-1}]m(\x,\theta,\mbP_0)| =O_p(s_{n,1}+s_{n,2}),
\ese
where the last step follows from the fact that $O_p(s_{n,1}+s_{n,2})=o(1)$ and repeated use of (\ref{num}), (\ref{deno}) as well as Assumption \ref{akernel} (ii). \hfill{$\square$}

\subsection{Proof of Proposition \ref{thbn}}\label{proof_thbn}
Considering that
\bse
\hphik(\X,\theta,\hmbP)\equiv \{\hl(\x,\hmbP)\}^{-1}\hm(\x,\theta,\hmbP)
\ese
with $\hm(\x,\theta,\mbP)\equiv h_n^{-r}\Enk[\{I(Y<\theta)-\tau\}K_h\{\mbP\trans(\x-\X)\}]$, it is obvious that, given $\cl$,
\bse
\{\hphik(\X,\theta,\hmbP):\theta\in\mbtv\}\subset\{\hphik(\X,\theta_i,\hmbP):i=1,\ldots,n+1\}
\ese
for any $\theta_1<Y_{(1)}$, $\theta_i\in[Y_{(i-1)},Y_{(i)})$ $(i=2,\ldots,n)$ and $\theta_{n+1}\geq Y_{(n)}$, where $Y_{(i)}$ is the $i$th order statistic of $\{Y_i:i=1,\ldots,n\}$. Therefore the set $\{\hphik(\X,\theta,\hmbP):\theta\in\mbtv\}$ contains at most $(n+1)$ different functions indexed by $\theta$ given $\cl$. This, combined with
(\ref{fstar}), implies that the set
$
\mp_{n,k}\equiv\{\hphik(\X,\theta)-\phi(\X,\theta):\theta\in\mbtv\}
$
satisfies\tcmAC{:}
\bse
N_{[\,]}\{\eta,\mp_{n,k}\mid\cl,L_2(\P_\X)\}\leq c\,(n+1)\eta^{-1}.  \qquad\qquad \hfill{\tcmAC{\mbox{$\square$}}}
\ese
%\hfill{$\square$}

\section{Additional simulation results}\label{sm_additional_simulations}
We %now
present \tcr{here} the results of \tcr{our} simulation \tcr{studies} %simulations
with $p=10$ or $200$\tcr{,} in Tables \ref{sm_table_simulation_efficiency} and \ref{sm_table_simulation_inference}. See Section \ref{secs} for the descriptions of the settings and \tcr{the} methods. \tcr{The behavior of the results -- both in estimation and inference -- are similar to the other cases presented in the main paper.}
%\abhishek{Why this bizarre formatting issue with the placement of tables??!! See if you can fix this. \tcb{It has been fixed.}}

\vspace{-0.1in}

\section{Supplement to the data analysis in Section \ref{secda}} \label{sm_data_analysis}
The following Table \ref{sm_table_data_analysis} lists the names and descriptions of the covariates \tcr{we considered for our data analysis in Section \ref{secda}. \tcm{These were the covariates included in all our imputation models.}} %used to construct our semi-supervised estimators.}
%included by the model of the data analysis in Section \ref{secda}.
\vspace{-0.15in}
\begin{table}[H]
	\def~{\hphantom{0}}
	\caption{Simulation results of Section \ref{secs}: Efficiencies of \tcr{the} semi-supervised estimators relative to the supervised estimator. The \textbf{boldface} in each case represents the best efficiency\tcr{.}}
	\vskip 2mm
	
	{
		\resizebox{\textwidth}{!}{
			\begin{tabular}{c|c|ccccc||cccc||cccc}
				\hline
				\hline
				&         & \multicolumn{5}{c||}{$p=10$}           & \multicolumn{4}{c||}{$p=200$, $q=5$} & \multicolumn{4}{c}{$p=200$,   $q=\ceil{p^{1/2}}$} \\
				$n$                   & $m(\X)$ & KS$_1$ & KS$_2$ & PR   & RF   & ORE  & KS$_1$   & KS$_2$  & PR    & ORE   & KS$_1$      & KS$_2$      & PR        & ORE       \\
				\hline
				\multirow{5}{*}{200}  & (a) & 0.94          & 0.91          & 0.93 & \textbf{0.94} & 1.00 & 0.96          & 0.81          & \textbf{0.98} & 1.00 & 0.96          & 0.81          & \textbf{0.98} & 1.00 \\
				& (b) & \textbf{3.42} & 3.25          & 3.07 & 1.63          & 3.33 & \textbf{2.54} & 2.29          & 2.18          & 2.52 & \textbf{3.24} & 2.22          & 1.81          & 3.90 \\
				& (c) & \textbf{2.98} & 2.88          & 2.84 & 1.55          & 3.21 & \textbf{2.02} & 1.80          & 1.58          & 2.23 & \textbf{2.62} & 2.13          & 1.81          & 3.86 \\
				& (d) & 2.51          & \textbf{2.66} & 2.34 & 1.54          & 2.88 & \textbf{1.45} & 1.25          & 1.42          & 2.14 & \textbf{2.20} & 1.77          & 1.65          & 3.41 \\
				& (e) & \textbf{2.20} & 2.16          & 1.94 & 1.56          & 3.58 & \textbf{1.80} & 1.66          & 1.58          & 2.67 & \textbf{1.75} & 1.43          & 1.44          & 4.19 \\
				%&     &               &               &      &               &      &               &               &               &      &               &               &               &      \\
				\hline
				\multirow{5}{*}{500}  & (a) & 0.98          & \textbf{0.99} & 0.96 & 0.96          & 1.00 & 0.97          & 0.93          & \textbf{0.99} & 1.00 & 0.97          & 0.93          & \textbf{0.99} & 1.00 \\
				& (b) & \textbf{3.19} & 3.12          & 3.10 & 1.75          & 2.95 & \textbf{1.98} & 1.90          & 1.92          & 2.33 & \textbf{3.19} & 3.00          & 2.61          & 3.37 \\
				& (c) & 2.86          & \textbf{2.87} & 2.85 & 1.67          & 2.86 & \textbf{1.94} & 1.86          & 1.65          & 2.09 & 2.94          & \textbf{3.02} & 2.59          & 3.34 \\
				& (d) & \textbf{2.52} & 2.51          & 2.37 & 1.63          & 2.62 & 1.39          & \textbf{1.41} & 1.38          & 2.01 & \textbf{2.40} & 2.26          & 2.04          & 3.01 \\
				& (e) & \textbf{2.13} & 2.08          & 2.12 & 1.74          & 3.14 & \textbf{2.01} & 1.94          & 1.88          & 2.45 & \textbf{2.24} & 2.12          & 2.00          & 3.57 \\
				%&     &               &               &      &               &      &               &               &               &      &               &               &               &      \\
				\hline
				\multirow{5}{*}{2000} & (a) & 0.98          & \textbf{1.01} & 0.99 & 0.97          & 1.00 & 0.99          & 1.00          & \textbf{1.00} & 1.00 & 0.99          & 1.00          & \textbf{1.00} & 1.00 \\
				& (b) & 1.95          & \textbf{1.96} & 1.94 & 1.58          & 2.08 & 1.89          & \textbf{1.89} & 1.85          & 1.81 & \textbf{2.19} & 2.17          & 2.10          & 2.23 \\
				& (c) & \textbf{1.93} & 1.92          & 1.91 & 1.54          & 2.05 & \textbf{1.68} & 1.67          & 1.51          & 1.70 & 2.11          & \textbf{2.11} & 2.03          & 2.22 \\
				& (d) & 1.74          & \textbf{1.76} & 1.70 & 1.47          & 1.94 & 1.46          & \textbf{1.61} & 1.43          & 1.65 & 2.09          & \textbf{2.13} & 1.99          & 2.10 \\
				& (e) & 1.67          & 1.68          & 1.66 & \textbf{1.69} & 2.15 & 1.71          & \textbf{1.72} & 1.69          & 1.87 & \textbf{1.61} & 1.60          & 1.57          & 2.30 \\
				\hline
	\end{tabular}}}
	\label{sm_table_simulation_efficiency}
	
	\begin{footnotesize}
		\tcr{\underline{Glossary of notation}:}	$p$, the dimension of $\X$; $q$, the sparsity level; $n$, the labeled data size; $m(\X)\equiv\E(Y\mid\X)$; KS$_1$/KS$_2$, kernel smoothing on the one/two direction(s) selected by linear regression/sliced inverse regression; RF, random forest; PR, parametric regression; ORE, oracle relative efficiency.
	\end{footnotesize}
\end{table}
%\vspace{-0.2in}

\begin{table}[H]
	\def~{\hphantom{0}}
	\caption{\tcb{Simulation results of Section \ref{secs}:} Inference based on  the semi-supervised estimators using kernel smoothing on the direction selected by linear regression. \underline{All the numbers have been multi-} \underline{plied by 100}. The \textbf{boldfaces} are the coverage rates of \tcr{the} 95\% confidence intervals\tcr{.}}
	\vskip 1mm
	
	{
		\resizebox{\textwidth}{!}{
			\begin{tabular}{c|c|cccc||cccc||cccc}
				\hline
				\hline
				&         & \multicolumn{4}{c||}{$p=10$} & \multicolumn{4}{c||}{$p=200$, $q=5$} & \multicolumn{4}{c}{$p=200$,   $q=\ceil{p^{1/2}}$} \\
				$n$                   & $m(\X)$ & ESE   & Bias & ASE  & CR   & ESE     & Bias   & ASE    & CR     & ESE        & Bias       & ASE        & CR         \\
				\hline
				\multirow{5}{*}{200}  & (a) & 9.0  & 0.3  & 9.7  & \textbf{97.2} & 8.8  & -0.2 & 9.6  & \textbf{96.6} & 8.8  & -0.2 & 9.6  & \textbf{96.6} \\
				& (b) & 17.3 & 0.1  & 17.6 & \textbf{93.4} & 13.3 & -0.1 & 15.1 & \textbf{97.4} & 19.6 & -1.7 & 20.8 & \textbf{95.8} \\
				& (c) & 17.7 & -0.6 & 16.7 & \textbf{92.4} & 13.0 & -1.4 & 15.1 & \textbf{97.6} & 21.6 & -1.7 & 21.9 & \textbf{95.6} \\
				& (d) & 16.8 & -0.4 & 18.3 & \textbf{96.6} & 14.9 & -0.3 & 18.6 & \textbf{98.4} & 20.7 & -1.1 & 23.4 & \textbf{96.4} \\
				& (e) & 23.0 & -0.1 & 23.5 & \textbf{95.0} & 17.2 & -0.8 & 18.1 & \textbf{96.0} & 28.3 & -1.1 & 30.6 & \textbf{95.0} \\
				%&     &      &      &      & \textbf{}     &      &      &      & \textbf{}     &      &      &      & \textbf{}     \\
				\hline
				\multirow{5}{*}{500}  & (a) & 5.8  & 0.0  & 5.9  & \textbf{95.6} & 5.9  & -0.1 & 5.9  & \textbf{95.4} & 5.9  & -0.1 & 5.9  & \textbf{95.4} \\
				& (b) & 10.9 & -0.4 & 11.4 & \textbf{97.4} & 9.7  & -0.4 & 9.5  & \textbf{93.8} & 12.5 & -0.5 & 13.1 & \textbf{96.6} \\
				& (c) & 11.0 & -0.8 & 10.6 & \textbf{95.2} & 8.4  & -0.8 & 9.0  & \textbf{96.4} & 12.9 & -0.7 & 12.8 & \textbf{95.6} \\
				& (d) & 10.3 & 0.5  & 11.3 & \textbf{96.8} & 9.6  & -0.4 & 11.0 & \textbf{97.4} & 12.5 & -0.6 & 13.6 & \textbf{97.4} \\
				& (e) & 14.6 & -1.0 & 14.9 & \textbf{93.8} & 10.4 & 0.4  & 11.2 & \textbf{97.2} & 15.9 & -0.6 & 18.1 & \textbf{97.6} \\
				%&     &      &      &      & \textbf{}     &      &      &      & \textbf{}     &      &      &      & \textbf{}     \\
				\hline
				\multirow{5}{*}{2000} & (a) & 2.8  & 0.3  & 2.9  & \textbf{96.2} & 2.6  & 0.3  & 2.9  & \textbf{96.6} & 2.6  & 0.3  & 2.9  & \textbf{96.6} \\
				& (b) & 6.6  & -0.2 & 6.6  & \textbf{94.0} & 5.2  & 0.0  & 5.3  & \textbf{95.4} & 7.5  & -1.1 & 7.7  & \textbf{95.4} \\
				& (c) & 6.2  & -0.3 & 6.2  & \textbf{95.6} & 4.8  & -0.1 & 4.7  & \textbf{95.2} & 7.6  & -1.0 & 7.5  & \textbf{94.2} \\
				& (d) & 5.9  & 0.0  & 6.2  & \textbf{96.2} & 5.1  & 0.1  & 5.5  & \textbf{96.2} & 6.8  & -0.8 & 7.4  & \textbf{97.6} \\
				& (e) & 7.2  & -1.2 & 8.0  & \textbf{97.0} & 5.8  & 0.1  & 6.0  & \textbf{95.2} & 9.2  & -1.1 & 9.7  & \textbf{96.8} \\
				\hline
	\end{tabular}}}
	\label{sm_table_simulation_inference}
	\vspace{-0.06in}
	
	\begin{footnotesize}
		\tcr{\underline{Glossary of notation}:} $p$, the dimension of $\X$; $q$, the sparsity level; $n$, the labeled data size; $m(\X)\equiv\E(Y\mid\X)$; ESE, empirical standard error; ASE, average of estimated standard errors; CR, coverage rate of 95\% confidence intervals.
	\end{footnotesize}
\end{table}
\vspace{-0.3in}
\begin{table}[H]
\def~{\hphantom{0}}
\caption{Covariates included \tcr{for the} %included by the model of the
	data analysis in Section \ref{secda}\tcr{.}}
\vskip 1mm

{
	\begin{tabular}{l|l}
		\hline
		\hline
		\textbf{Variable   name} & \textbf{Description}                                           \\
		\hline
		active &
		In your usual day, how active are you in 1971? \\
		age                      & Age in 1971                                                    \\
		alcoholfreq &
		How often do you drink in 1971?    \\
		allergies & Use allergies medication in 1971 \\
		asthma &  DX asthma in 1971 \\
		cholesterol              & Serum cholesterol (mg/100ml) in 1971                           \\
		dbp                      & Diastolic blood pressure in 1982                               \\
		education &
		Amount of education by 1971 \\
		exercise &
		In recreation, how much exercise in 1971? \\
		ht                       & Height in centimeters in 1971                                  \\
		price71                  & Average tobacco price in state of residence   1971 (US\$2008)      \\
		price82                  & Average tobacco price in state of residence 1982 (US\$2008)        \\
		race                     & White, black or other in 1971                             \\
		sbp                      & Systolic blood pressure in 1982                                \\
		sex                      & Male or female                                              \\
		smokeintensity           & Number of cigarettes smoked per day in 1971                    \\
		smokeyrs                 & Years of smoking                                               \\
		tax71                    & Tobacco tax in state of residence 1971 (US\$2008)              \\
		tax82                    & Tobacco tax in state of residence 1971 (US\$2008)              \\
		wt71                     & Weight in kilograms  in 1971                                   \\
		\hline
\end{tabular}}
\label{sm_table_data_analysis}
\end{table}
	
\end{appendix}

\phantomsection
\addcontentsline{toc}{section}{References}

\bibliographystyle{apalike}
%\bibliography{IntegralApprox}

\bibliography{myreference-ssq}  %% Needed to create this alternative .bib file to clean up some MathSci based citations that were becoming a nuisance under JMLR style citations.%%

\end{document}